\documentclass[twocolumn]{aastex62}

\usepackage{longtable}
\usepackage{comment}
\usepackage{color}
\usepackage{hyperref}
\usepackage{soul}

\newcommand{\cxc}{{\em Chandra }}

\newcommand{\red}{\textcolor{red}}

\shorttitle{{\em CHANDRA} OBSERVATIONS OF A3411--12}
\shortauthors{Andrade-Santos et al.}

\begin{document}

%% LaTeX will automatically break titles if they run longer than
%% one line. However, you may use \\ to force a line break if
%% you desire.

%%\title{{\em Chandra} observations of the spectacular A3411--12 merger event}
\title{{\em CHANDRA} OBSERVATIONS OF THE SPECTACULAR A3411--12 MERGER EVENT}

%% Use \author, \affil, and the \and command to format
%% author and affiliation information.
%% Note that \email has replaced the old \authoremail command
%% from AASTeX v4.0. You can use \email to mark an email address
%% anywhere in the paper, not just in the front matter.
%% As in the title, use \\ to force line breaks.

\author{Felipe Andrade-Santos}
\affil{Clay Center Observatory, Dexter Southfield, 20 Newton Street, Brookline, MA 02445, USA}
\affil{Center for Astrophysics \text{\textbar} Harvard \& Smithsonian , 60 Garden Street, Cambridge, MA 02138, USA}
\email{fsantos@cfa.harvard.edu}

\author{Reinout J. van Weeren}
\affil{Leiden Observatory, Leiden University, PO Box 9513, 2300 RA Leiden, The Netherlands}

\author{Gabriella Di Gennaro}
\affil{Leiden Observatory, Leiden University, PO Box 9513, 2300 RA Leiden, The Netherlands}

\author{David Wittman}
\affil{Department of Physics, University of California, Davis, CA 95616, USA}

\author{Dongsu Ryu}
\affil{Department of Physics, School of Natural Sciences, UNIST, Ulsan 44919, Republic of Korea}

\author{Dharam Vir Lal}
\affil{National Centre for Radio Astrophysics - Tata Institute of Fundamental Research, Post Box 3, Ganeshkhind P.O., Pune 41007, India}

\author{Vinicius M. Placco}
\affil{Department of Physics, University of Notre Dame, Notre Dame, IN 46556, USA}
\affil{Joint Institute for Nuclear Astrophysics, Center for the Evolution of the Elements, East Lansing, MI 48824, USA}

\author{Kevin Fogarty}
\affil{Division of Physics, Math, and Astronomy, California Institute of Technology, Pasadena, CA, USA}
\affil{Space Telescope Science Institute, Baltimore, MD, USA}

\author{M. James Jee}
\affil{Yonsei University, Department of Astronomy, Seoul, Republic of Korea}
\affil{Department of Physics, University of California, Davis, CA 95616, USA}

\author{Andra Stroe}
\thanks{Clay Fellow}
\affil{Center for Astrophysics \text{\textbar} Harvard \& Smithsonian , 60 Garden Street, Cambridge, MA 02138, USA}

\author{David Sobral}
\affil{Department of Physics, Lancaster University, Lancaster, LA1 4YB, UK}
\affil{Leiden Observatory, Leiden University, PO Box 9513, 2300 RA Leiden, The Netherlands}

\author{William R. Forman}
\affil{Center for Astrophysics \text{\textbar} Harvard \& Smithsonian , 60 Garden Street, Cambridge, MA 02138, USA}

\author{Christine Jones}
\affil{Center for Astrophysics \text{\textbar} Harvard \& Smithsonian , 60 Garden Street, Cambridge, MA 02138, USA}

\author{Ralph P. Kraft}
\affil{Center for Astrophysics \text{\textbar} Harvard \& Smithsonian , 60 Garden Street, Cambridge, MA 02138, USA}

\author{Stephen S. Murray}
\affil{Department of Physics and Astronomy, The Johns Hopkins University, 3400 N. Charles St., Baltimore, MD 21218, USA}

\author{Marcus Br\"uggen}
\affil{Hamburger Sternwarte, University of Hamburg, Gojenbergsweg 112, D-21029 Hamburg, Germany}

\author{Hyesung Kang}
\affil{Department of Earth Sciences, Pusan National University, Busan 46241, Republic of Korea}

\author{Rafael Santucci}
\affil{Instituto de Estudos S\'ocio-Ambientais, Planet\'ario, Universidade Federal de Goi\'as, Goi\^ania, GO 74055-140, Brazil}
\affil{Instituto de F\'isica, Universidade Federal de Goi\'as, Campus Samambaia, Goi\^ania, GO 74001-970, Brazil}

\author{Nathan Golovich}
\affil{Lawrence Livermore National Laboratory, 7000 East Avenue, Livermore, CA 94550, USA}

\author{William Dawson}
\affil{Lawrence Livermore National Laboratory, 7000 East Avenue, Livermore, CA 94550, USA}

%% Notice that each of these authors has alternate affiliations, which
%% are identified by the \altaffilmark after each name.  Specify alternate
%% affiliation information with \altaffiltext, with one command per each
%% affiliation.

%% Mark off your abstract in the ``abstract'' environment. In the manuscript
%% style, abstract will output a Received/Accepted line after the
%% title and affiliation information. No date will appear since the author
%% does not have this information. The dates will be filled in by the
%% editorial office after submission.

%%%%%%%%%%%%%%%%%%%%%%%%%%%%%%%%%%%%%%%%%%%%%%%%%%%%%%%%%%%%%%%%%%%%%%%%%%%%%%%%%%
%%%%%%%%%%%%%%%%%%%%%%%%%%%%%%%%%%%%%%%%%%%%%%%%%%%%%%%%%%%%%%%%%%%%%%%%%%%%%%%%%%
%%
%%                                  ABSTRACT
%%
%%%%%%%%%%%%%%%%%%%%%%%%%%%%%%%%%%%%%%%%%%%%%%%%%%%%%%%%%%%%%%%%%%%%%%%%%%%%%%%%%%
%%%%%%%%%%%%%%%%%%%%%%%%%%%%%%%%%%%%%%%%%%%%%%%%%%%%%%%%%%%%%%%%%%%%%%%%%%%%%%%%%%

\begin{abstract}

We present deep {\em Chandra} observations of 
A3411--12, a remarkable merging cluster that hosts the most compelling evidence 
for electron re-acceleration at cluster shocks to date. 
Using the $Y_X-M$ scaling relation, we find $r_{500} \sim 1.3$ Mpc, $M_{500}
= (7.1 \pm 0.7) \times
10^{14} \ M_{\rm{\odot}}$, $kT=6.5\pm 0.1$ keV, and a gas mass of
$M_{\rm g,500} = (9.7 \pm 0.1) \times 10^{13}
M_\odot$.
The gas mass fraction within $r_{500}$ is 
$f_{\rm g} = 0.14 \pm 0.01$.
We compute the shock strength 
using density jumps to conclude that the Mach number of the merging subcluster is small ($M \leq 1.15_{-0.09}^{+0.14}$). 
We also present pseudo-density, projected temperature, pseudo-pressure, and
pseudo-entropy maps. Based on the pseudo-entropy map we conclude that the cluster is undergoing a mild merger,
consistent with the small Mach number.
On the other hand, radio relics extend 
over Mpc scale in the A3411--12 system, which strongly suggests that a population of energetic electrons already existed over extended regions of the cluster. 

\end{abstract}

%% Keywords should appear after the \end{abstract} command. The uncommented
%% example has been keyed in ApJ style. See the instructions to authors
%% for the journal to which you are submitting your paper to determine
%% what keyword punctuation is appropriate.


\section{Introduction}

Galaxy cluster mergers are the most energetic events
in the present day Universe and involve kinetic energies
on the order of $\sim 10^{63-64}$ erg. Direct evidence
for cluster mergers has been found from the
morphology of the X-ray emission \citep[e.g.,][]{1984Jones,1999Jones,1995Mohr,1996Buote,2005Jeltema,2010Lagana,
2012Andrade-Santos,2013Andrade-Santos} and the presence
of shocks \citep[e.g.,][]{2002Markevitch} in the intracluster
medium (ICM), as well as from the asymmetric
spatial and velocity distributions of cluster galaxy
populations \citep{1988Dressler}.

The identification and study of merging clusters
is of considerable astrophysical interest for several
reasons. First, such major mergers are rare events,
and have a profound, long lasting influence on the
thermodynamic evolution of the ICM. Major mergers
are believed to be responsible for the general division
of clusters into cool-core and non-cool-core
clusters \citep{2009Henning}. Mergers can also affect
a wide range of other cluster related phenomena,
including AGN activity in cluster galaxies \citep{2010Ma,2015Sobral}
and star formation \citep{2008Lagana,2015Sobral,2015Stroe,2017Stroe}. Second, cluster mergers
are an ideal laboratory to study the properties of dark
matter. X-ray and optical (lensing) studies have put
strong constraints on the self-interaction of dark
matter and have shown that it must be nearly collisionless
\citep{2004Clowe,2006Clowe,2006Bradac,2008Bradac,2008Randall,2012Dawson}.

\begin{figure*}[hbt!]
\centerline{%
\includegraphics[width=0.98\textwidth]{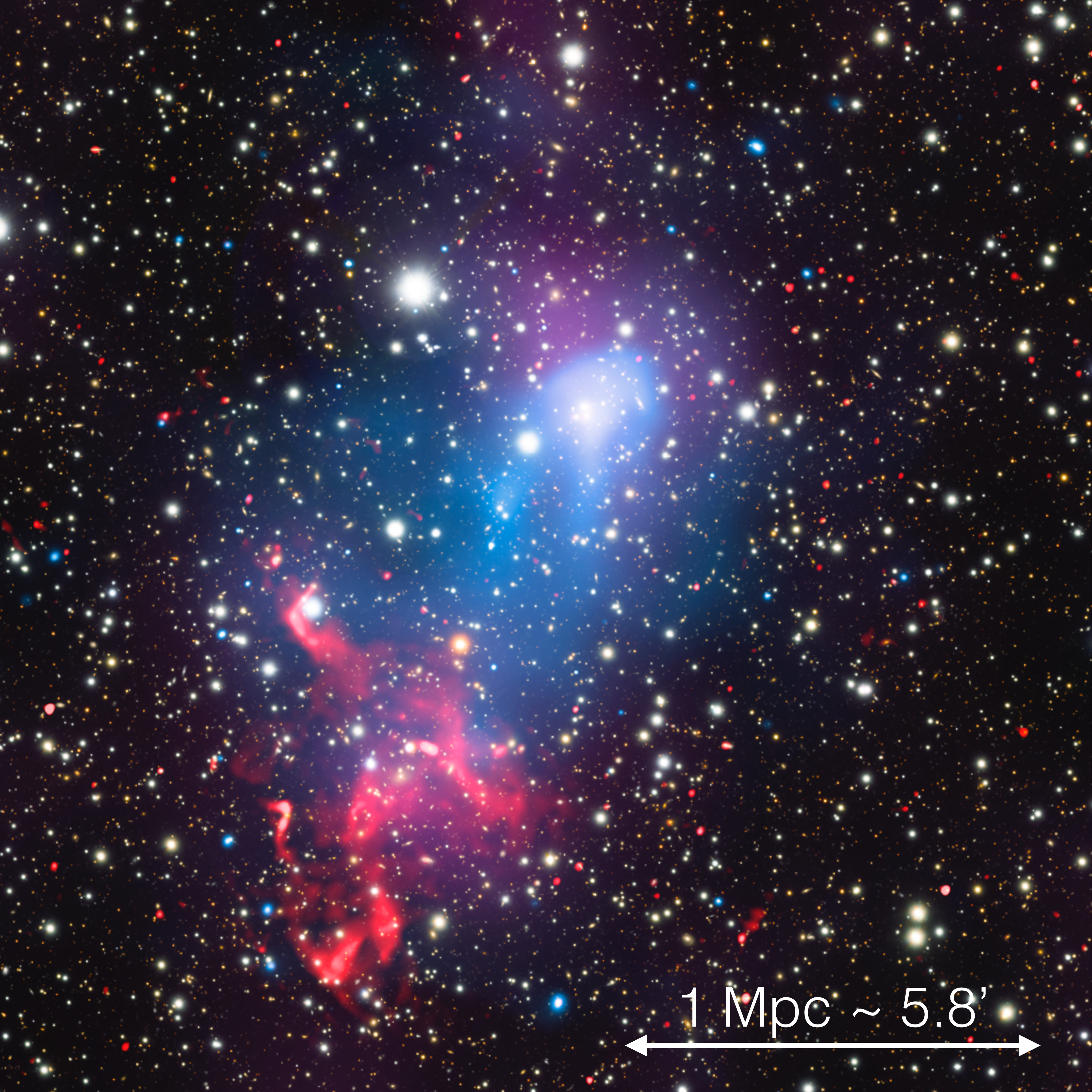}
}
\caption{
\small{
Composite image of the A3411--A3412 field: optical (Subaru -- RGB), 0.5-2.0 keV X-ray ({\em Chandra} -- in
blue), 325 and 610 MHz radio (GMRT -- in red), and the galaxy density distribution (purple). A3411--12 presents a clear cool core in the north and a
large (Mpc scale) radio relic in the south. In the north there is an
overdensity of galaxies at the cluster redshift, located in the cool
core (X-ray: bright blue, galaxy density: purple) region, whereas another peak in the galaxy
distribution (purple) is seen in the relic region (red) in the south.
}
}
\label{fig:optical_xray_radio}
\end{figure*}

% INTRO MERGERS, LSS
Simulations of large-scale structure formation show that galaxy clusters 
grow through gas accretion from large-scale filaments and mergers of smaller 
clusters and groups. 
These mergers are characterized by the enormous amounts of energy involved 
($\sim 10^{64}$~erg), long lifetimes (Gyr), and large physical scales (Mpc). 
Most of the gravitational energy released during a merger event is converted 
to thermal energy via shocks and turbulence \citep[see][for a review]{2007Markevitch}. In addition, a small fraction ($< 1$\%) of the shock energy 
could be channeled into the acceleration of cosmic rays (CR). In the presence 
of magnetic fields these CR then emit synchrotron radiation, which can be 
observed with radio telescopes. 

% INTRO RELICS and the low-Mach problem
The  elongated and arc-like radio sources that trace cluster merger shocks are commonly called radio relics \citep[][see Figure 1]{2012Feretti}.
A major problem in our understanding is how these low-Mach number ($\mathcal{M}$) 
cluster merger shocks can accelerate enough particles to explain {\red the} observed radio synchrotron brightness. 
According to  standard diffusive shock acceleration theory \citep[DSA;][]{1983Drury},
the acceleration efficiency is very low for $\mathcal{M} \lesssim 4$ shocks, and the existence of 
radio relics is therefore very puzzling \citep[e.g.,][]{2014Hong,2012Kang}. 

Two main particle acceleration mechanisms have been proposed to explain radio relics:

% 2. SHOCK ACCELERATION
(1) Shock acceleration: Particles gain energy via multiple
crossing of the shock front, however, it is very hard to reconcile the low
acceleration efficiency with the bright radio relics. 

% 1. RE-ACCELERATION
(2) Re-acceleration: Shocks re-accelerate a population of pre-existing (``fossil'') relativistic electrons via DSA 
\citep{2005Markevitch}, avoiding the low acceleration efficiency problem. 
Good source candidates for these fossil electrons are radio
galaxies (common in clusters). 

\subsection{A3411--12}

\begin{figure}[t!]
\centerline{%
\includegraphics[width=0.47\textwidth]{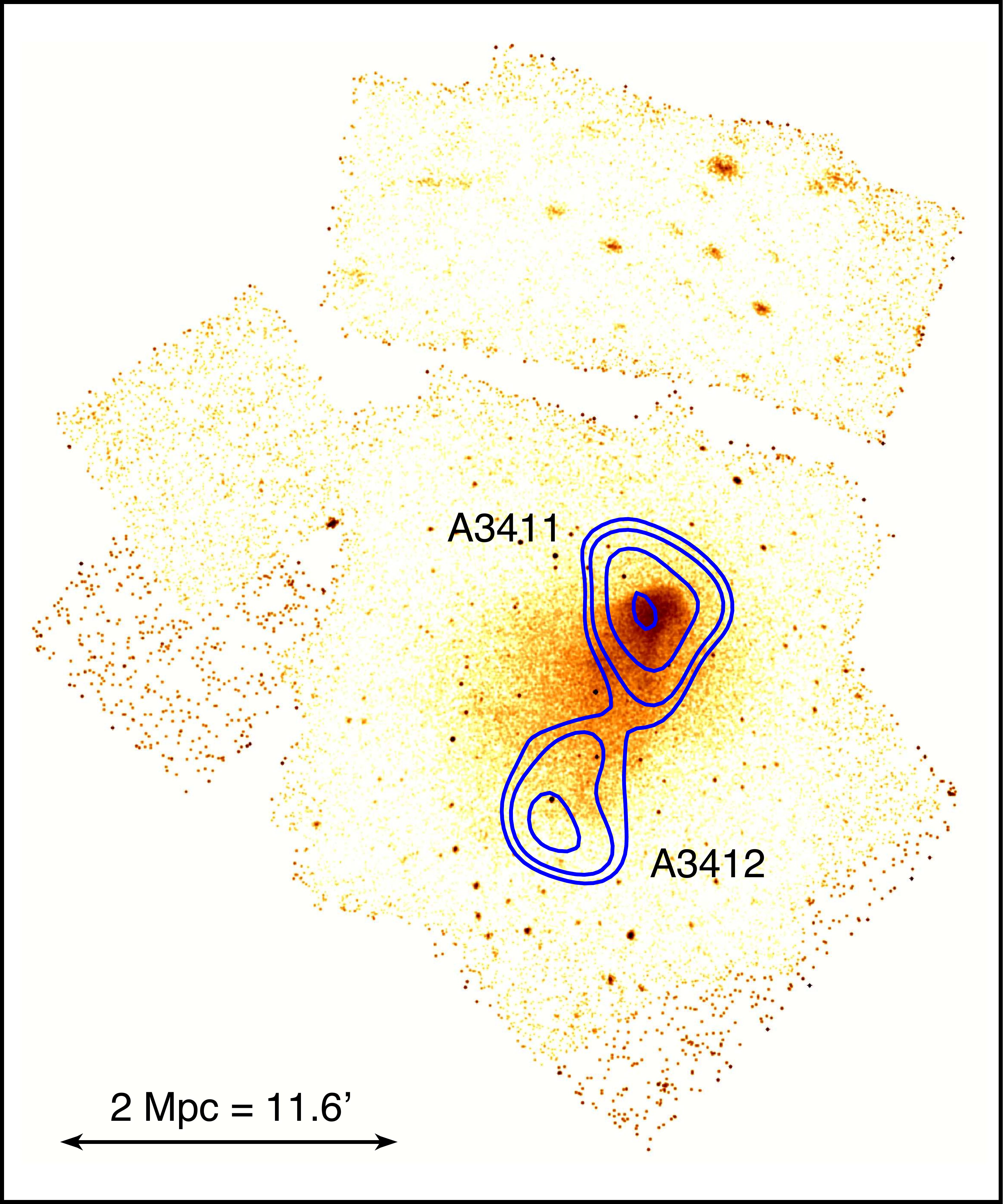}
}
\caption{
\small{
0.5--2.0 keV, background subtracted, flat-field {\em Chandra} image of the
A3411--12 field overlaid by galaxy iso-density contours in blue.
}
}
\label{fig:mosaic}
\end{figure}

A3411--12 (also known as PLCKESZ G241.97+14.85 -- see Figure \ref{fig:optical_xray_radio}) is a relatively nearby ($z =
0.1687$) merging cluster presenting a large ($\sim0.7$ Mpc) radio
relic \citep{2013VanWeeren,2013Giovannini}. 
\citet{2013VanWeeren} showed that A3411--12 is a merging system, 
with the projected merger axis oriented SE-NW (see
Figure \ref{fig:optical_xray_radio}) with A3411 in the NW and A3412 in the SE. {\em Chandra} X-ray image shows that the cluster has a cometary shape with a well-defined
subcluster core visible in the northwestern part of the system (A3411).
Fainter X-ray emission is found surrounding the subcluster core
and this emission seems to be part of a second, larger subcluster (A3412). 
There is no clear surface brightness peak corresponding
to the primary cluster core (A3412 -- see Figure \ref{fig:mosaic}), which suggests the primary cluster
has been disrupted by the collision with the subcluster (A3411 -- see Figure \ref{fig:mosaic}), as
has been the case for A2146 \citep{2011Russell,2012Menanteau}.
\citet{2013Giovannini} have also found that the radio halo at the center of A3411-12 has a low power 
(${\rm log}~ P = 23.16  {\rm ~ W ~ Hz}^{-1}$).
The global ICM
temperature of A3411--12 is $\sim$ 6 keV and its X-ray 0.5--2.0 keV luminosity is
$2.8 \times 10^{44} \rm~erg~s^{-1}$ within $r_{500}$ = 1.34 Mpc \citep{2013VanWeeren}. 

More recently,  \citet{2017vanWeerenNatAst} found, in the merging galaxy cluster A3411--12, the most compelling evidence for re-acceleration at cluster shocks to date.
The authors identified a tailed radio galaxy connected to the relic.
In addition, spectral flattening is observed at the location where the fossil  
plasma meets the relic and, at the same location, an X-ray surface brightness edge is observed.

\citet{2017vanWeerenNatAst} also presented a clustering analysis applied to the three-dimensional galaxy distribution (right ascension, declination, and redshift) of their spectroscopic sample of cluster members (obtained with Keck). They considered mixtures of 1 to 7 multivariate Gaussian components. They found that, of the models considered, the two-component Gaussian model is the most favored one, indicating a bi-modal distribution.  They also investigated the redshift and velocity dispersion of each subcluster. They found similar velocity dispersions for the northern (A3411) and southern (A3412) subclusters, $1110^{+100}_{-80}\rm  ~km~s^{-1}$
and $1190^{+100}_{-90} \rm ~km~s^{-1}$, respectively. These velocity dispersions translated into mass
estimates of $1.4^{+0.4}_{-0.3} \times 10^{15} M_\odot$ and $1.8^{+0.5}_{-0.4} \times 10^{15} M_\odot$ for the A3411 and A3412 subclusters, respectively. With the mass estimates and redshift distributions, they concluded that core passage for the A3411-A3412 merger event occurred about $\sim$ 1 Gyr before the photons {\em Chandra} collected were emitted and that the plane of the merger event is seen relatively close ($9^\circ$--$41^\circ$) to the plane of the sky crossing the cluster center, implying that the shock is seen close to edge-on. It is worthy mentioning that in the current work we find a much smaller total mass for the system of $M_{\rm 500} = (7.14 \pm 0.65) \times 10^{14} M_\odot$, in very good agreement with the {\em Planck} estimated mass of $M_{\rm 500} = (6.59 \pm 0.31) \times 10^{14} M_\odot$ \citep{2016PlanckCol}.

In this {\it paper} we characterize this cluster based on Chandra X-ray observations, present 
temperature, density, pressure, and entropy maps, as well as density
jumps related to cold and shock fronts.
We show that if indeed the density jumps trace shocks, they are mild, 
indicating that a population of energetic electrons already existed 
over extended regions of the cluster based on the extension of the radio relics in the A3411--12
system.
The cosmology assumed for our analysis has $\Omega_{\rm M}=0.3$,
$\Omega_{\Lambda}=0.7$ and $H_0=70$~km~s$^{-1}$Mpc$^{-1}$, implying a
linear scale of $2.88\rm ~ kpc ~ arcsec^{-1}$ at the A3411--12 luminosity distance of 812 Mpc ($z=0.1687$). All uncertainties are 68\% confidence level, unless otherwise stated.

%%%%%%%%%%%%%%%%%%%%%%%%%%%%%%%%%%%%%%%%%%%%%%%%%%%%%%%%%%%%%%%%%%%%%%%%%%%%%%%%%%
%%%%%%%%%%%%%%%%%%%%%%%%%%%%%%%%%%%%%%%%%%%%%%%%%%%%%%%%%%%%%%%%%%%%%%%%%%%%%%%%%%
%%
%%                   CHANDRA OBSERVATION AND DATA REDUCTION          
%%
%%%%%%%%%%%%%%%%%%%%%%%%%%%%%%%%%%%%%%%%%%%%%%%%%%%%%%%%%%%%%%%%%%%%%%%%%%%%%%%%%%
%%%%%%%%%%%%%%%%%%%%%%%%%%%%%%%%%%%%%%%%%%%%%%%%%%%%%%%%%%%%%%%%%%%%%%%%%%%%%%%%%%

\section{X-ray observations and data reduction}

We observed A3411 with the \cxc ~X-ray Observatory
(ACIS-I detectors, VF mode, ObsIds 
%% Dataset identifiers, as requested by CXC.  Keep.
%\dataset[ADS/Sa.CXO#obs/13378]{13378} and
%\dataset[ADS/Sa.CXO\#obs/15316]{15316} -- PI Murray, and \dataset[ADS/Sa.CXO\#obs/17193]{17193},
%\dataset[ADS/Sa.CXO\#obs/17496]{17496},\dataset[ADS/Sa.CXO\#obs/17497]{17497},\dataset[ADS/Sa.CXO\#obs/17583]{17583},
%\dataset[ADS/Sa.CXO\#obs/17585]{17585}, and
%\dataset[ADS/Sa.CXO\#obs/17584]{17584} -- PI van Weeren). 
13378, 15316 -- PI Murray, and 17193, 17496, 17583, 17585, 17584 -- PI van Weeren).
The data were reduced using the software \texttt{CHAV} which follows the processing described in
\citet{2005Vik}, applying the calibration files \texttt{CALDB 4.6.7}. 
The data processing includes corrections for the time dependence of the charge
transfer inefficiency and gain, and a check for periods
of high background (none were found -- the total exposure time is 211 ks).
Also, readout artifacts were
subtracted and standard blank sky background files were used for 
background subtraction. Figure \ref{fig:mosaic} shows the combined image
of all observations in the 0.5--2.0 keV energy band. 

Since the focus of this paper is solely on X-ray data and their results, we refer the reader to
\citet{2017vanWeerenNatAst} for the details on the optical and radio reductions
performed to create the image displayed on Figure \ref{fig:optical_xray_radio}.

%%%%%%%%%%%%%%%%%%%%%%%%%%%%
%%%%%%%%%%%%%%%%%%%%%%%%%%%%
%%
%%                     Overall Characteristics 
%%
%%%%%%%%%%%%%%%%%%%%%%%%%%%%
%%%%%%%%%%%%%%%%%%%%%%%%%%%%

\section{Overall characteristics of the cluster}\label{sec:gas_properties}

%%%%%%%%%%%%%%%%%%%%%%%%%%%%
%%%%%%%%%%%%%%%%%%%%%%%%%%%%
%%
%%                      EMM PROFILE
%%
%%%%%%%%%%%%%%%%%%%%%%%%%%%%
%%%%%%%%%%%%%%%%%%%%%%%%%%%%

\subsection{Emission measure profile}\label{section:emmprof} 

In this section we outline the procedures used to compute the emission measure profile. We refer the reader to \citet{2006Vik} for a detailed description of the method.

\begin{figure*}[!t]
\centerline{
\vspace{-0.5cm}
\includegraphics[width=0.47\textwidth, bb= -290 -500 1230 1000]{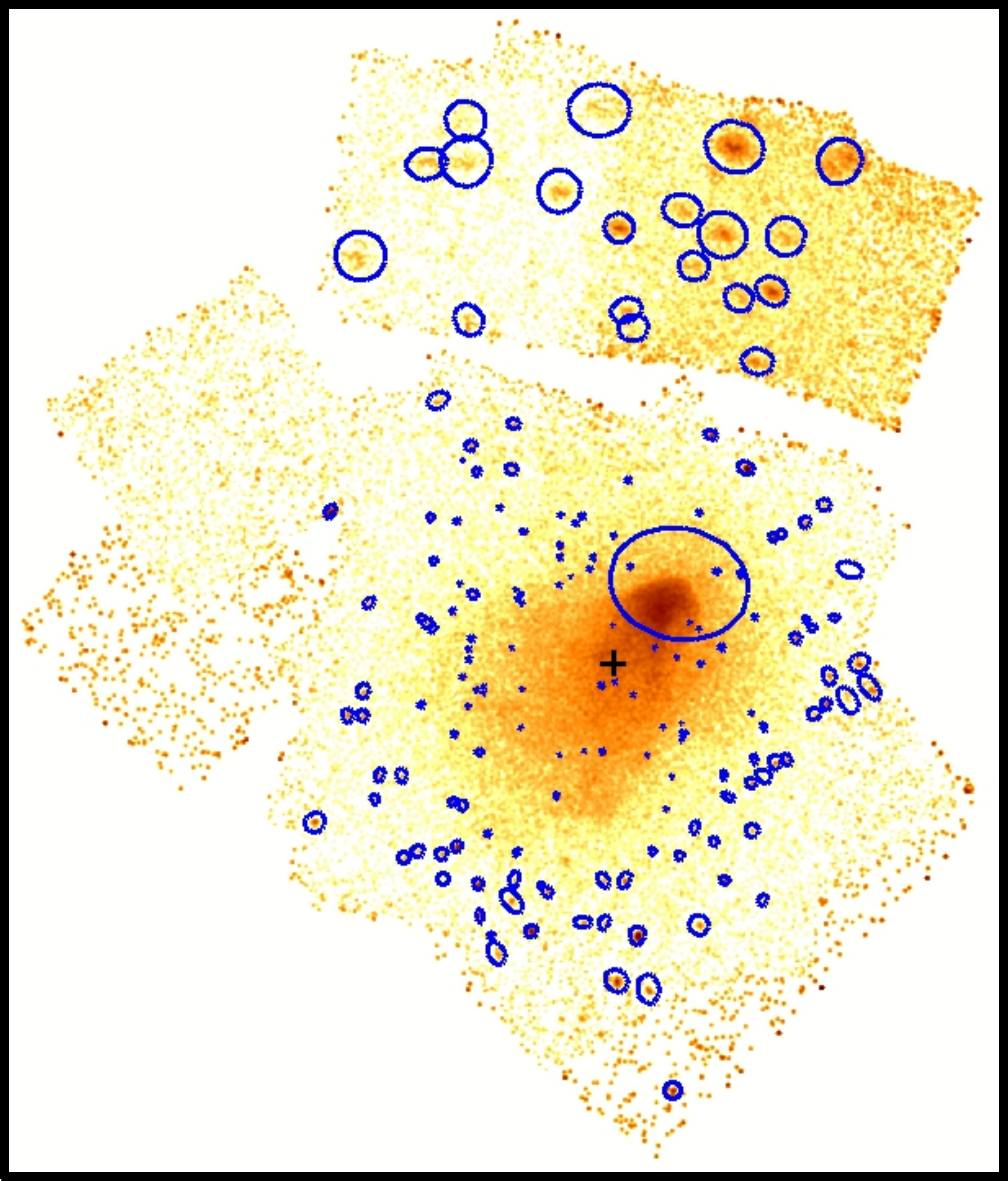}
\includegraphics[width=0.47\textwidth]{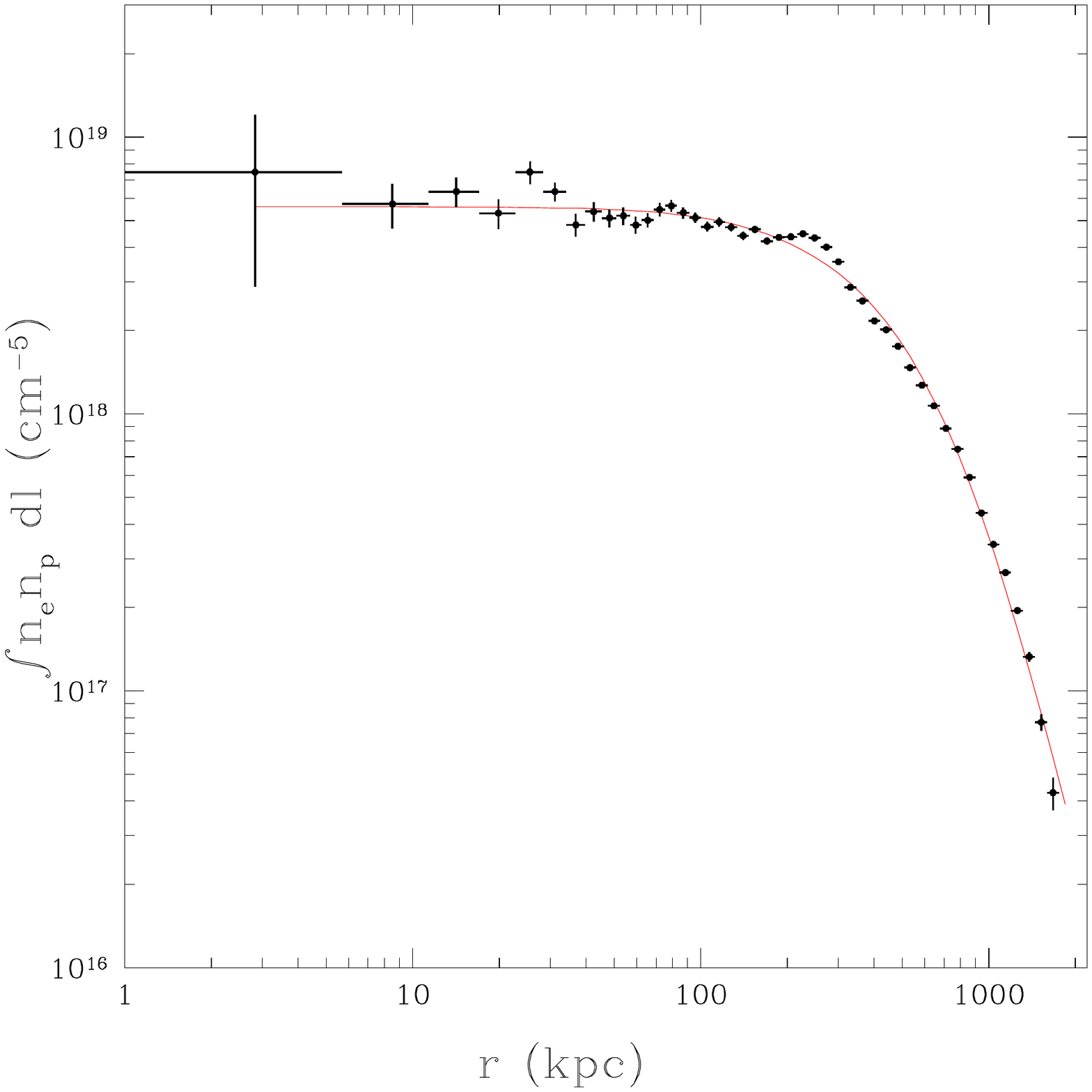}
\vspace{-3cm}
}
\centerline{
\includegraphics[width=0.47\textwidth]{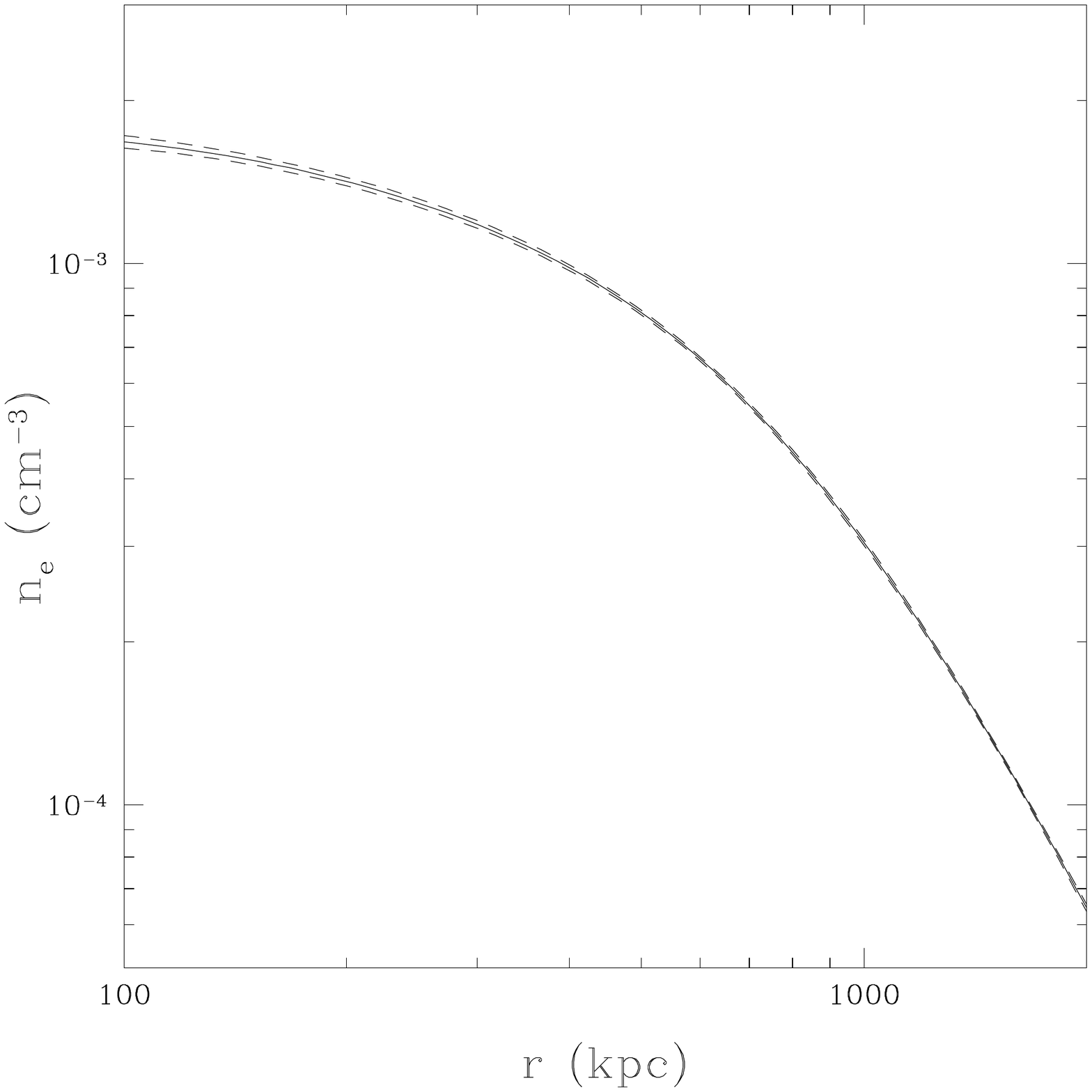}
\includegraphics[width=0.47\textwidth]{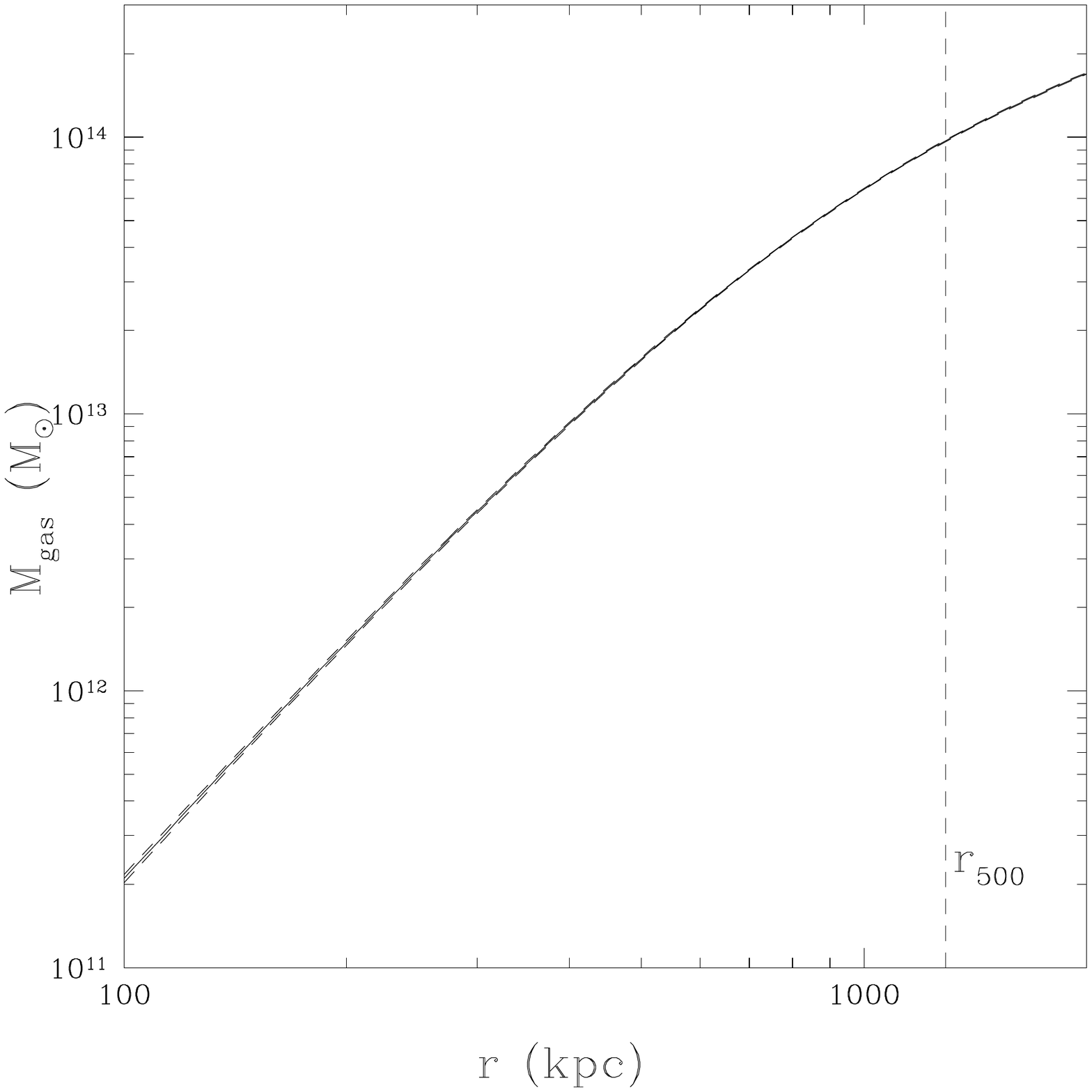}
\vspace{-1.5cm}
}
\caption{\small{
X-ray image (upper left), projected
emissivity (upper right), gas density (lower left),
and enclosed gas mass (lower right) profiles for A3411--12.
Top left panel shows the 0.5-2.0 keV, background-subtracted, exposure map corrected
ACIS-I image. The total filtered {\em Chandra} exposure is 211
ks.
Blue ellipses correspond to the masked X-ray point
sources (we also masked the bullet (cool-core in A3411))
and the black cross corresponds to the cluster center 
(determined by computing the X-ray centroid in a circle of $\sim$ 1.3 Mpc radius including the cool-core).
Top right panel shows the projected emissivity profile. The
solid line shows the emission measure integral of the best fit
to the emissivity profile given by Equation (\ref{eq:nenp}) assuming 
$n_{\rm e} = 1.1995 \times n_{\rm p}$, where $n_{\rm e}$ and $n_{\rm p}$ are the electron and proton number densities, respectively. Bottom left panel shows the electron number density profile. The
solid line shows the electron number density profile obtained from the emissivity
profile given by Equation
(\ref{eq:nenp}).
Bottom right panel shows the gas mass profile, with the dashed
vertical line indicating $r_{500}$. The dashed lines in the
electron number density and gas mass profiles show the 68\%
confidence range.
}}\label{fig:emm_dens}
\end{figure*}

First we detected compact sources using \texttt{wavdect} in the 0.7--2.0 keV or 2.0--7.0 keV bands and then masked these from the spectral and spatial analyses (we also masked the bullet (cool-core in A3411) -- see top left panel of Figure \ref{fig:emm_dens}). We then measured the surface brightness profiles in the 0.7--2.0 keV energy band, which maximizes the signal to noise ratio in {\em Chandra} data. The readout artifacts and blank-field background \citep[see section 2.3.3 of][]{2006Vik} were subtracted from the X-ray images, and the result was exposure-corrected using exposure maps
(computed assuming an absorbed optically-thin thermal plasma with $kT = 5.0$
keV, abundance = 0.3 $Z_{\sun}$, plus the Galactic column density\footnote{NH was fixed to the Galactic value, taking into account not
only the 21 cm map of the Galactic atomic hydrogen but also the molecular
contribution ($\rm NH_{\rm total} = NHI + NH_2 =
(4.67+1.25) \times 10^{20} = 5.92 \times 10^{20}$ --
\href{http://www.swift.ac.uk/analysis/nhtot/}{http://www.swift.ac.uk/analysis/nhtot/}).}
that include corrections for bad pixels and CCD gaps, but do not 
take into account spatial variations of the effective area. 
Finally, we subtracted any small uniform component corresponding to 
soft X-ray foreground adjustments that may be required.

\begin{deluxetable*}{cccccccccc}[!t]
\tablecaption{Parameters for the emission measure profile (Equation \ref{eq:nenp})} 
\tablewidth{0pt} 
\tablehead{ 
\colhead{$n_0$} &
\colhead{$r_{\rm c}$} & 
\colhead{$r_{\rm s}$} &
\colhead{$\alpha$} &
\colhead{$\beta$} &
\colhead{$\gamma$} & 
\colhead{$\epsilon$} &
\colhead{$n_{02}$} &
\colhead{$r_{\rm c2}$} & 
\colhead{$\beta_2$} \\
\colhead{($\rm 10^{-3} ~ cm^{-3}$)} &
\colhead{(kpc)} &
\colhead{(kpc)} &
\colhead{} &
\colhead{} &
\colhead{} &
\colhead{} &
\colhead{($\rm 10^{-3} ~ cm^{-3}$)} &
\colhead{(kpc)} &
\colhead{} 
}
\startdata 
$1.12 \pm 0.19$ & $260 \pm 3$ & $262 \pm 9$ & $0.00 \pm 0.10$ & $0.75 \pm 0.20$ & $0.79 \pm 0.11$ & $0.04 \pm 0.10$ & $1.23 \pm 0.07$ & $736 \pm 52$ & $0.96 \pm 0.06$
\enddata
\tablecomments{Columns list best fit values for the parameters
given by Equation \ref{eq:nenp}.} 
\label{tab:emm_prof}
\end{deluxetable*}

\begin{deluxetable*}{cccccccc}[!t]
\tablecaption{Parameters for the Temperature Profile (Equations \ref{eq:tprof:main} and \ref{eq:tprof:cool})} 
\tablewidth{0pt} 
\tablehead{ 
\colhead{$T_0$} &
\colhead{$T_{\rm min}$} & 
\colhead{$r_{\rm t}$} &
\colhead{$r_{\rm cool}$} &
\colhead{$a_{\rm cool}^\dag$} &
\colhead{$a^\dag$} & 
\colhead{$b$} &
\colhead{$c$} \\
\colhead{(keV)} &
\colhead{(keV)} &
\colhead{(kpc)} &
\colhead{(kpc)} &
\colhead{} &
\colhead{} &
\colhead{} &
\colhead{} 
}
\startdata 
$13.0 \pm 1.7$ & $5.2 \pm 0.7$ & $545 \pm 173$ & $400 \pm 104$ & 1.9 & 0 & $5.0 \pm 1.3$ & $3.1 \pm 1.9$ 
\enddata
\tablecomments{Columns list best fit values for the parameters
given by Equations \ref{eq:tprof:main} and \ref{eq:tprof:cool}. $^\dag$ fixed value.} 
\label{tab:temp_prof}
\end{deluxetable*}

Following these steps, we extracted the surface brightness profiles in narrow 
concentric annuli ($r_{\rm out}/r_{\rm in} = 1.05$) centered on the
X-ray centroid (determined excluding the masked regions) and computed the {\em Chandra} area-averaged effective area for each
annulus \citep[see][for details on calculating the effective area]{2005Vik}. 
Using the observed projected temperature, effective area, and 
metallicity as a function of radius, we  converted the {\em Chandra} count 
rate in the 0.7--2.0 keV band into the emission integral, 
${\rm EI} =  \int n_{\rm e} n_{\rm p} dV$, within each cylindrical
shell. The X-ray morphology of A3411--12 exhibits an irregular shape (see Figure \ref{fig:mosaic}), however, this is mostly due to the bullet in the northern part of the cluster. When 
we mask the bullet, the cluster exhibits a more elongated and symmetrical shape.
To compute the emission measure and temperature profiles we assumed spherical 
symmetry. In this case, the spherical assumption is expected to have only small effects on the total mass of the cluster when using $Y_{\rm X}$ as a proxy, as  presented by \citet{2006Kravtsov}. They showed that $Y_{\rm X}$ is a robust mass indicator with remarkably low scatter of only $\approx 5\%$--$7\%$ in $M_{500}$
for fixed $Y_{\rm X}$, regardless of whether the cluster is relaxed or not.
We then fit the emission measure profile assuming the gas density
profile follows \citet{2006Vik}: 

\begin{eqnarray}
n_{\rm e}n_{\rm p} &=& n_0^2
\frac{ (r/r_{\rm c})^{-\alpha}}{(1+r^2/r_{\rm c}^2)^{3\beta-\alpha/2}}
\frac{1}{(1+r^\gamma/r_{\rm s}^\gamma)^{\epsilon/\gamma}}+ \nonumber \\
&&\frac{n^2_{02}}{(1+r^2/r_{\rm c2}^2)^{3\beta_2}}.\label{eq:nenp}
\end{eqnarray} 

This relation is based on a classic $\beta$-model, modified to account
for the power-law type cusp and the steeper emission measure slope at large radii. In addition, 
a second $\beta$-model is included, giving extra freedom to characterize the cluster core. 
For further details on this equation we refer to \citet{2006Vik}. The
relation between the electron number density and gas mass density is given by $\rho_{\rm g} =
\mu_{\rm e} n_{\rm e} m_{\rm a}$, where $m_{\rm a}$ is the atomic mass
unit and $\mu_{\rm e}$ is the mean
molecular weight per electron. For a typical metallicity of
0.3 $Z_\odot$, the reference values from \citet{1989AndersGrevesse}
yield $\mu_{\rm e} = 1.17058$ and $n_{\rm e}/n_{\rm p} = 1.1995$, where $n_{\rm p}$ is the proton number density.
The best fit parameters of Equation \ref{eq:nenp} are listed in Table \ref{tab:emm_prof}. 
Figure \ref{fig:emm_dens} presents the best fit emission measure profile, as well as the density
and gas mass profiles derived from the best fit emission measure. The gas mass profile is then used 
to compute the total mass using the $Y_{\rm x}$ relation (Section \ref{sec:masses}). 

%%%%%%%%%%%%%%%%%%%%%%%%%%%%
%%%%%%%%%%%%%%%%%%%%%%%%%%%%
%%
%%                  TEMPERATURE PROFILE
%%
%%%%%%%%%%%%%%%%%%%%%%%%%%%%
%%%%%%%%%%%%%%%%%%%%%%%%%%%%

\subsection{Gas Temperature Radial Profiles}\label{sec:temp}

\begin{figure}[t!]
 \vspace{-1.2cm}
  \begin{center}
    \leavevmode
      \includegraphics[width=0.47\textwidth]{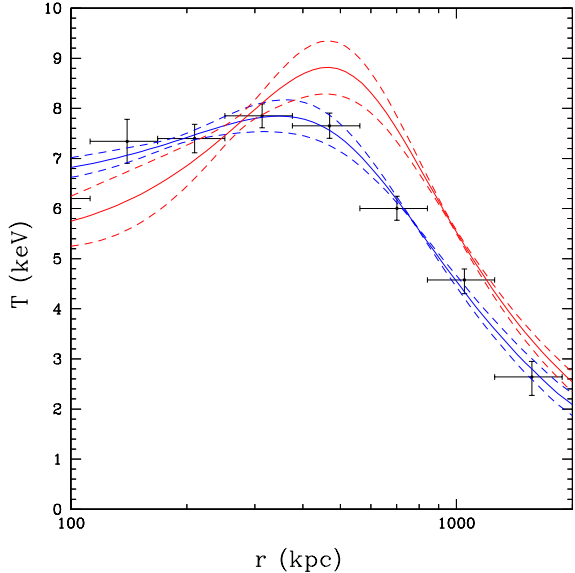}
      \vspace{-2.0cm}
      \caption{Azimuthally averaged, radial temperature profile. Observed
  projected temperatures are shown by points with error
  bars. The 3D model and its projected
effective temperatures (the latter to be compared with the data) are shown by the
red and blue curves, respectively. Dashed lines show the 1$\sigma$ uncertainty ranges.}\label{fig:temp_prof}
  \end{center}
\end{figure}

Most clusters present a temperature profile that has
a broad peak within 0.1--0.2 $r_{200}$\footnote{$r_{200}$ and
$r_{500}$ are used to define a radius at the over-density of 200 and
500 times the critical density of the Universe at the cluster
redshift, respectively.}. 
\citet{2006Vik} present a 3D temperature
profile that describes these general
features. At large radii, the temperature profile
can be reasonably well represented as a broken power law with a transition
region:
\begin{equation}\label{eq:tprof:main}
  T(r) =\frac{(r/r_t)^{-a}}{(1+(r/r_t)^b)^{c/b}}. 
\end{equation}

At small radii, the temperature profile can be described as
\begin{equation}\label{eq:tprof:cool}
  T_{\text{cool}}(r) = (x+T_{\text{min}}/T_0)/(x+1),
\end{equation}
where $x=(r/r_{\text{cool}})^{a_{\text{cool}}}$. The final analytical expression for the 3D temperature
profile is,
\begin{equation}\label{eq:tprof}
  T_{\mathrm{3D}}(r) =  T_0\times T_{\text{cool}}(r)\times T(r). 
\end{equation}

This temperature model has significant functional freedom (8 parameters) and can
adequately describe almost any smooth temperature distribution. 
Thus, we use this model, from \citet{2006Vik}, to describe the temperature distribution of the hot gas in A3411--12. 

To construct the temperature profile, we extracted spectra from 7 annuli
in the radial range from $\sim 100$ to
$\sim 2000$ kpc (logarithmically spaced in distance)
and fit them with an absorbed \texttt{APEC} model.
For the fitting we fixed NH to the Galactic value of $5.92 \times 10^{20}$ (see Section \ref{section:emmprof}).
We then followed the procedures described above to obtain the 2D and
3D temperature profiles. 
The measured 2D (black data points), fitted 2D (blue solid line), and
3D (red solid line) temperature profiles are presented
in Figure \ref{fig:temp_prof}. The 2D temperature
profile was computed by projecting the 3D temperature\red{,} weighted by gas density squared
using the spectroscopic-like temperature \citep[][provides a formula for the
temperature which matches the spectroscopically
measured temperature within a few percent]{2004Mazzotta}:
\begin{equation}
T_{\rm 2D}=T_{\rm spec} \equiv \frac{\int \rho_{\rm g}^2 T_{\rm
    3D}^{1/4} dz}{\int \rho_{\rm g}^2 T_{\rm 3D}^{-3/4} dz}
\label{eq:tspec}
\end{equation}

To estimate the uncertainties
in the best values for the parameters of this analytical model, we
performed Monte-Carlo simulations. 
This model for $T_{\mathrm{3D}}(r)$ (Equation \ref{eq:tprof}) allows
very steep temperature gradients. In some Monte-Carlo realizations,
such profiles are mathematically consistent with the observed projected
temperatures. However, large
values of temperature gradients often lead to unphysical mass
estimates, such as profiles with negative dark matter density at some
radii. We solved this issue by accepting only Monte-Carlo realizations in which the
best-fit temperature profile leads to $\rho_{\text{tot}}>\rho_{\text{gas}}$ in the radial
range $r \leq 1.5 r_{500}$, where $\rho_{\text{tot}} = \rho_{\rm gas} + \rho_{\rm dark~ matter}$.
Also, in the same radial range, we
verified that the temperature profiles are all convectively stable, i.e. $d\ln  T/d\ln  \rho_g < 2/3$. 

The best fit parameters of equations \ref{eq:tprof:main} and
\ref{eq:tprof:cool} are presented in Table \ref{tab:temp_prof}.

Interestingly, the temperature profile of A3411--12 is very smooth when the bullet is removed from the analysis (see Figure \ref{fig:temp_prof}), despite that this system is undergoing a major merger. 

%%%%%%%%%%%%%%%%%%%%%%%%%%%%
%%%%%%%%%%%%%%%%%%%%%%%%%%%%
%%
%%                     MASSES
%%
%%%%%%%%%%%%%%%%%%%%%%%%%%%%
%%%%%%%%%%%%%%%%%%%%%%%%%%%%

%%%%%%%%%%%%%%%%%%%%%%%%%%%%%%%%%%%%%%%%%%%%%%%%%%%%%%%%%%%%%%%%%%%%%%%%%%%%%%%%%%
%%%%%%%%%%%%%%%%%%%%%%%%%%%%%%%%%%%%%%%%%%%%%%%%%%%%%%%%%%%%%%%%%%%%%%%%%%%%%%%%%%
%%
%%                                  MASS ESTIMATES
%%
%%%%%%%%%%%%%%%%%%%%%%%%%%%%%%%%%%%%%%%%%%%%%%%%%%%%%%%%%%%%%%%%%%%%%%%%%%%%%%%%%%
%%%%%%%%%%%%%%%%%%%%%%%%%%%%%%%%%%%%%%%%%%%%%%%%%%%%%%%%%%%%%%%%%%%%%%%%%%%%%%%%%%

\section{Cluster Mass Estimates}\label{sec:masses}

Using the gas mass and temperature, we estimated
the total cluster mass from the $Y_X$--$M$ scaling relation of \citet{2009Vik},
\begin{eqnarray}
M_{\rm 500,Y_X} = E^{-2/5}(z)A_{\rm YM}\left(\frac{Y_{\rm X}}{3\times10^{14}M_\odot
  {\rm keV}}\right)^{B_{\rm YM}}, \nonumber \\
\label{e_yx_m}
\end{eqnarray}
where $Y_{\rm X} = M_{\rm gas,500} \times kT_{\rm X}$, $M_{\rm
  gas,500}$ is computed using the best fit parameters of Equation (\ref{eq:nenp}), and $T_{\rm X}$ is the measured
temperature in the (0.15--1) $\times ~
r_{500}$ range. 
$A_{\rm YM}=(5.77 \pm 0.20)\times10^{14}h^{1/2}  M_\odot$ and $B_{\rm YM}=0.57
\pm 0.03$ \citep{2009Vik}. Here,
$M_{\rm Y_X,500}$ is the total mass within $r_{500}$,
and $E(z)=[\Omega_{\rm M}(1+z)^3 + (1-\Omega_{\rm M}-\Omega_\Lambda)(1+z)^2 +
\Omega_\Lambda]^{1/2}$ is the function describing the evolution of the Hubble
parameter with redshift. 

Using Equation (\ref{e_yx_m}), $r_{500}$  is computed by
solving 
\begin{eqnarray}
M_{\rm 500,Y_X} \equiv 500 \rho_c (4\pi/3) r_{500}^{3},\label{m500_def}
\end{eqnarray}
where $\rho_c$ is the critical density of the Universe at the cluster
redshift. In practice, Equation (6) is evaluated at 
a given radius, whose result is compared
to the evaluation of Equation (7) at the same radius. This
process is repeated in an iterative procedure, until the fractional mass difference
is less than 1\%. We estimated 1$\sigma$ uncertainties in the $Y_{\rm X}$ derived masses 
using Monte Carlo simulations. We also added to the Monte Carlo
procedure a 1$\sigma$ systematic uncertainty of 9\% in the mass determination, as discussed by \citet{2009Vik}.

\begin{figure*}[t!]
\centerline{%
\includegraphics[width=0.98\textwidth]{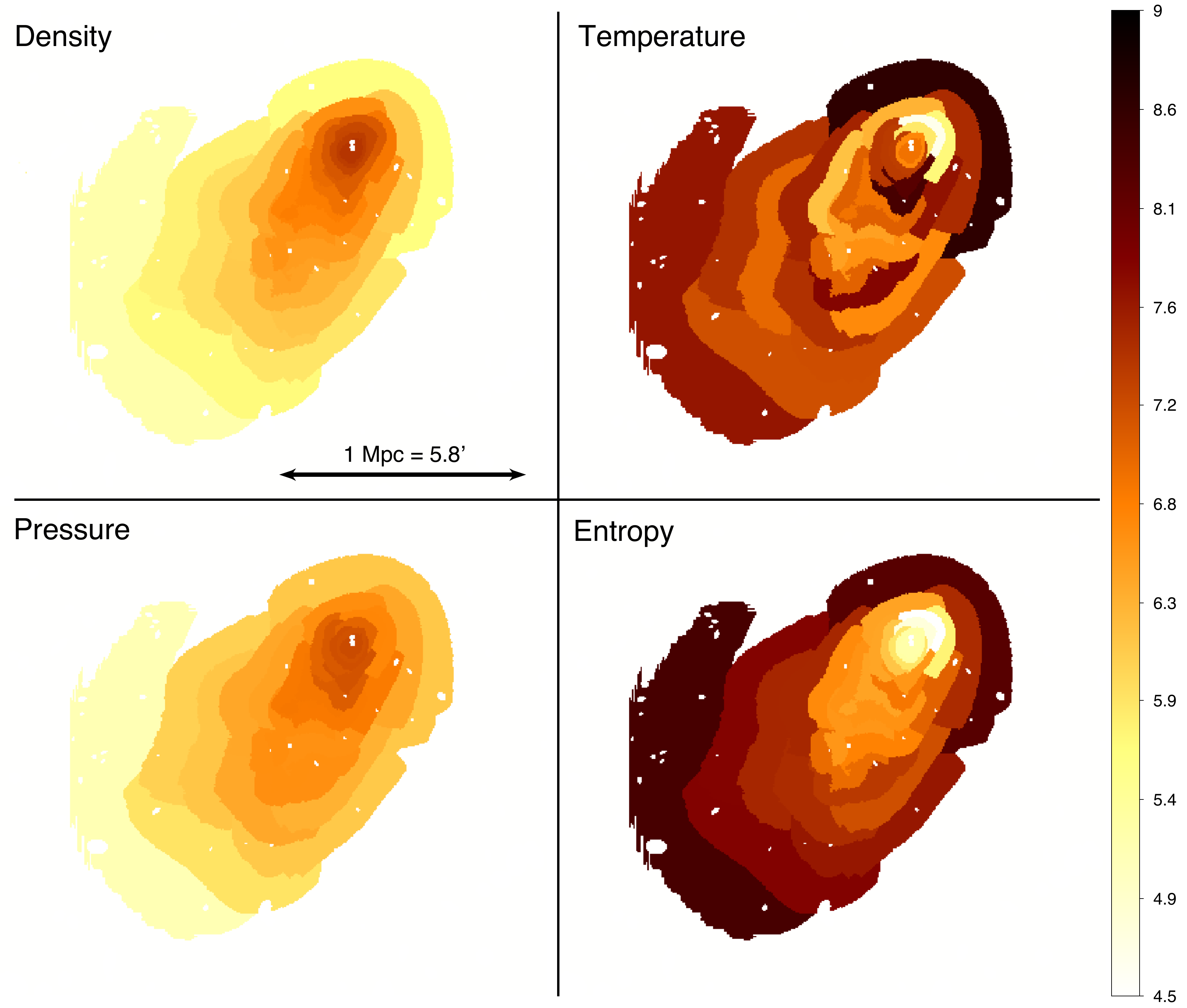}
}
\caption{
\small{
Top left: pseudo-density. Top right: projected temperature. Bottom left:
pseudo-pressure ($P = kT \times n_{\rm e}$). Bottom right: pseudo-entropy ($K = kT /
n_{\rm e}^{2/3}$). Color bar indicates the temperature. All maps are in
logarithmic scale and darker colors represent higher values. The white ellipses
represent the excluded point source regions. These figures are described in detail in the text.
}
}
\label{fig:rho_kT_P_K}
\end{figure*}

\begin{figure}[t!]
\centerline{%
\includegraphics[width=0.47\textwidth,angle=0]{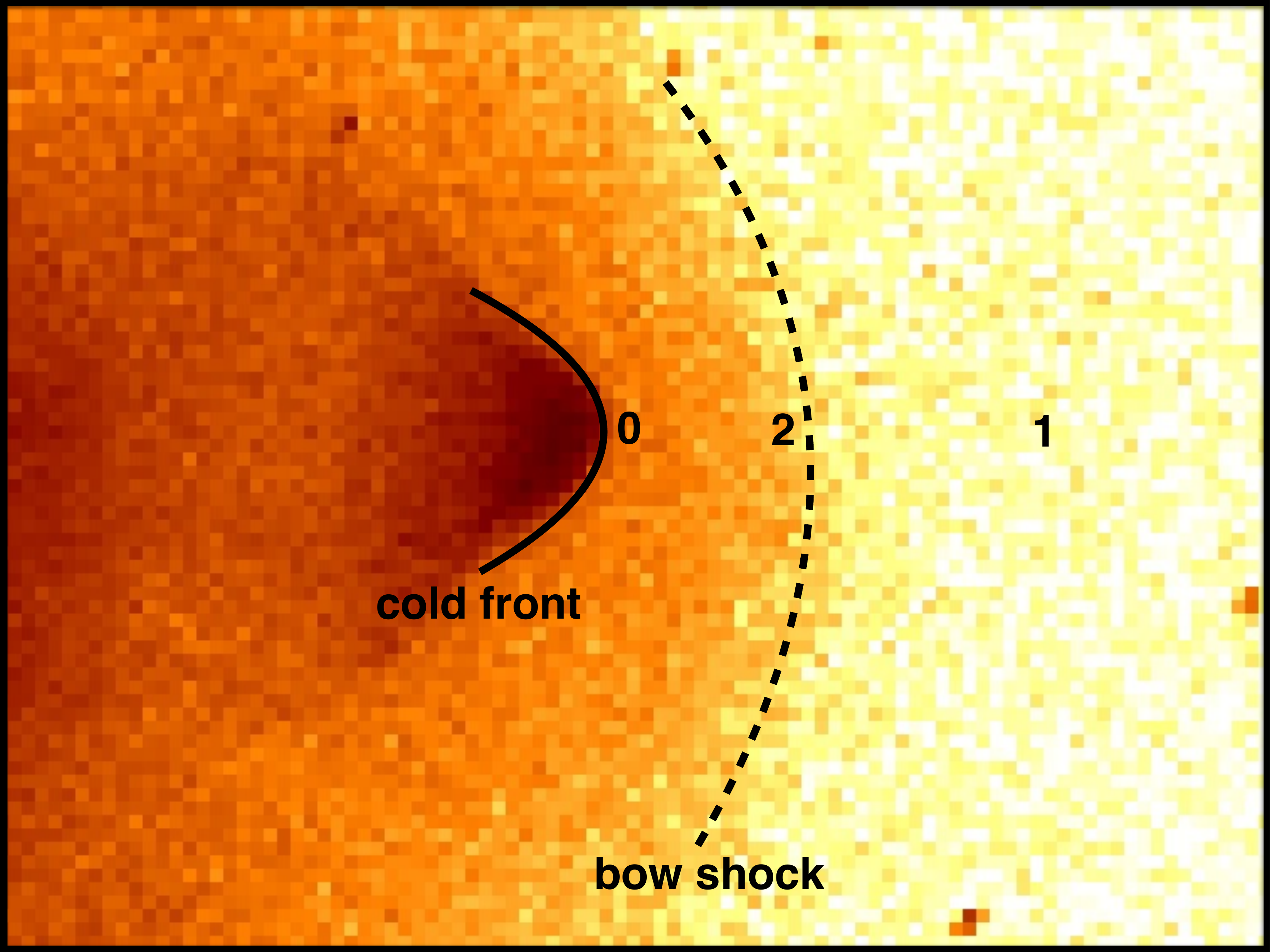}
}
\caption{
\small{
Geometry of flow past a denser and colder region.  Here the Bullet cluster is used as a textbook example of cold front and bow shock formations.
Zones 0, 1, and 2 are those near the stagnation point, in the undisturbed free stream, and past the bow shock, respectively. While cold fronts may be the result of many different
physical events in the cluster, the density discontinuities 
in them form for the same basic reason: whenever 
a gas density peak encounters a flow of ambient gas, a contact
discontinuity quickly forms.
}
}
\label{fig:geometry_of_flow}
\end{figure}

\section{Results}

\subsection{Masses}

Following the approach outlined in Section \ref{sec:masses}, we obtain  $M_{\rm 500,Y_X} = (7.14 \pm 0.65) \times 10^{14} M_\odot$, in very good agreement with the {\em Planck}
estimated mass of $M_{\rm 500,Y_{SZ}} = (6.59 \pm 0.31) \times 10^{14} M_\odot$ \citep{2016PlanckCol} despite the merger morphology. This is due to the fact that both $Y_{\rm X}$ and $Y_{\rm SZ}$ are insensitive to the cluster dynamical state \citep{2006Kravtsov,2013Sayers}. This mass leads to $r_{500} \sim 1.3$ Mpc.  We measure $kT = 6.5 \pm 0.1$ keV within (0.15--1.0) $\times ~r_{500}$, and a gas mass of
$M_{\rm g,500} = (9.7 \pm 0.1) \times 10^{13} M_\odot$. The gas mass fraction within $r_{500}$ is $f_{\rm g} = 0.14 \pm 0.01$. 

\subsection{Images}
\label{sec:images}
Figure \ref{fig:mosaic} shows the merged, flat-fielded 
(vignetting and exposure corrected), and background subtracted $0.5-2$ keV band 
\textit{Chandra} ACIS-I image of A3411--12. The image reveals the presence of large scale 
diffuse emission, which originates from optically-thin thermal plasma with $kT \sim$  2--8
keV temperature (Section \ref{sec:gas_properties}).

The distribution of the hot X-ray emitting gas reveals a complex
morphology, indicating an active merger history \citep{2013VanWeeren,2013Giovannini,2017vanWeerenNatAst}. In particular, the gas distribution 
is not symmetric, but is elongated in the southeast-northwest direction. In addition, the image 
shows the presence of sharp surface brightness edges in the central regions of the 
cluster (see Figure \ref{fig:regions_sb}).

The above features are characteristic signatures of a merger, which has likely 
perturbed the hot gas distribution. To explore the nature of these features, and hence, 
constrain the merger history of the cluster, we derive surface brightness, density, and 
temperature profiles, which are discussed in the following sections.

\begin{figure*}
\centering
{\includegraphics[width=0.49\textwidth]{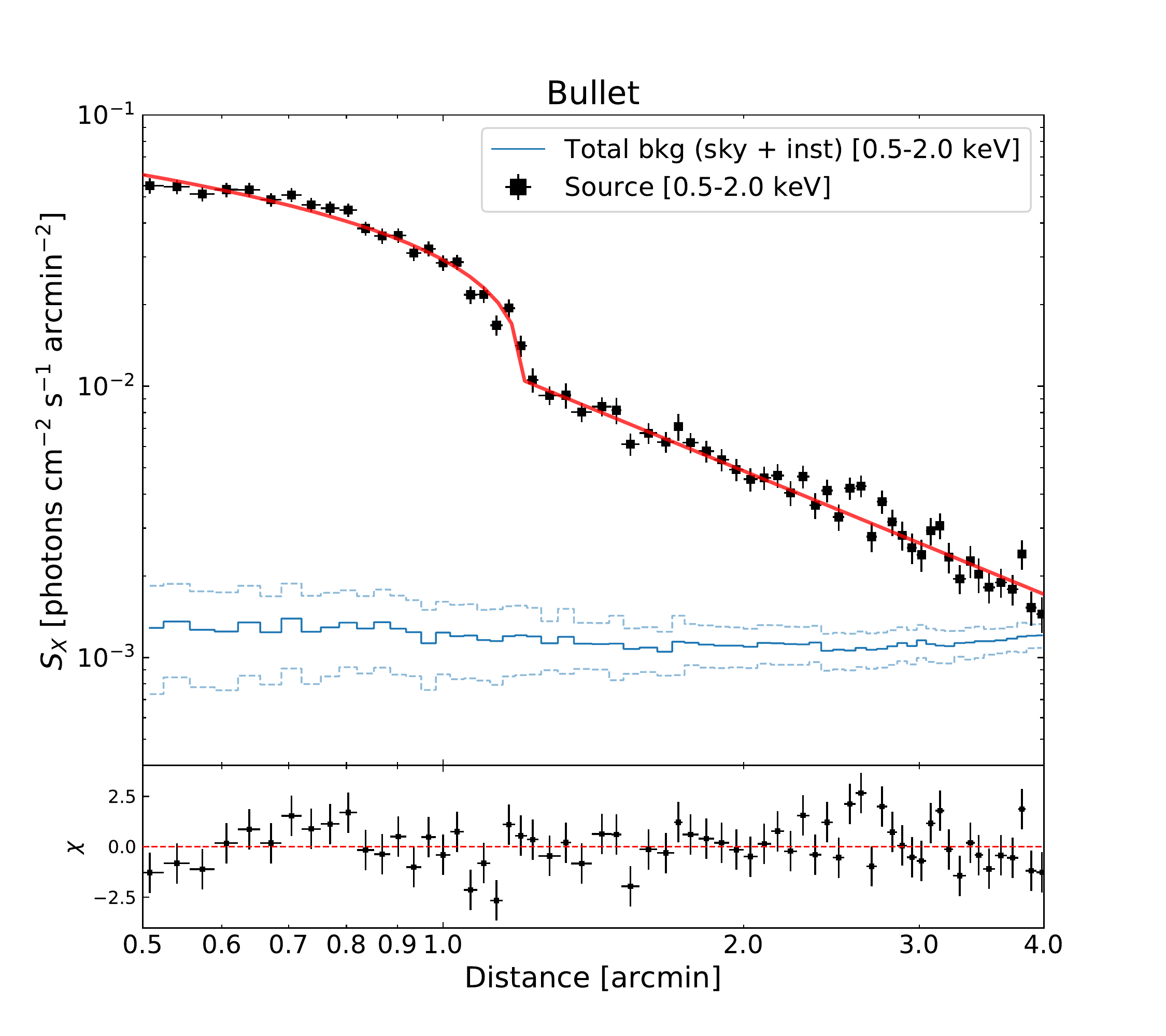}}
\hspace{0.16cm}
{\includegraphics[width=0.49\textwidth]{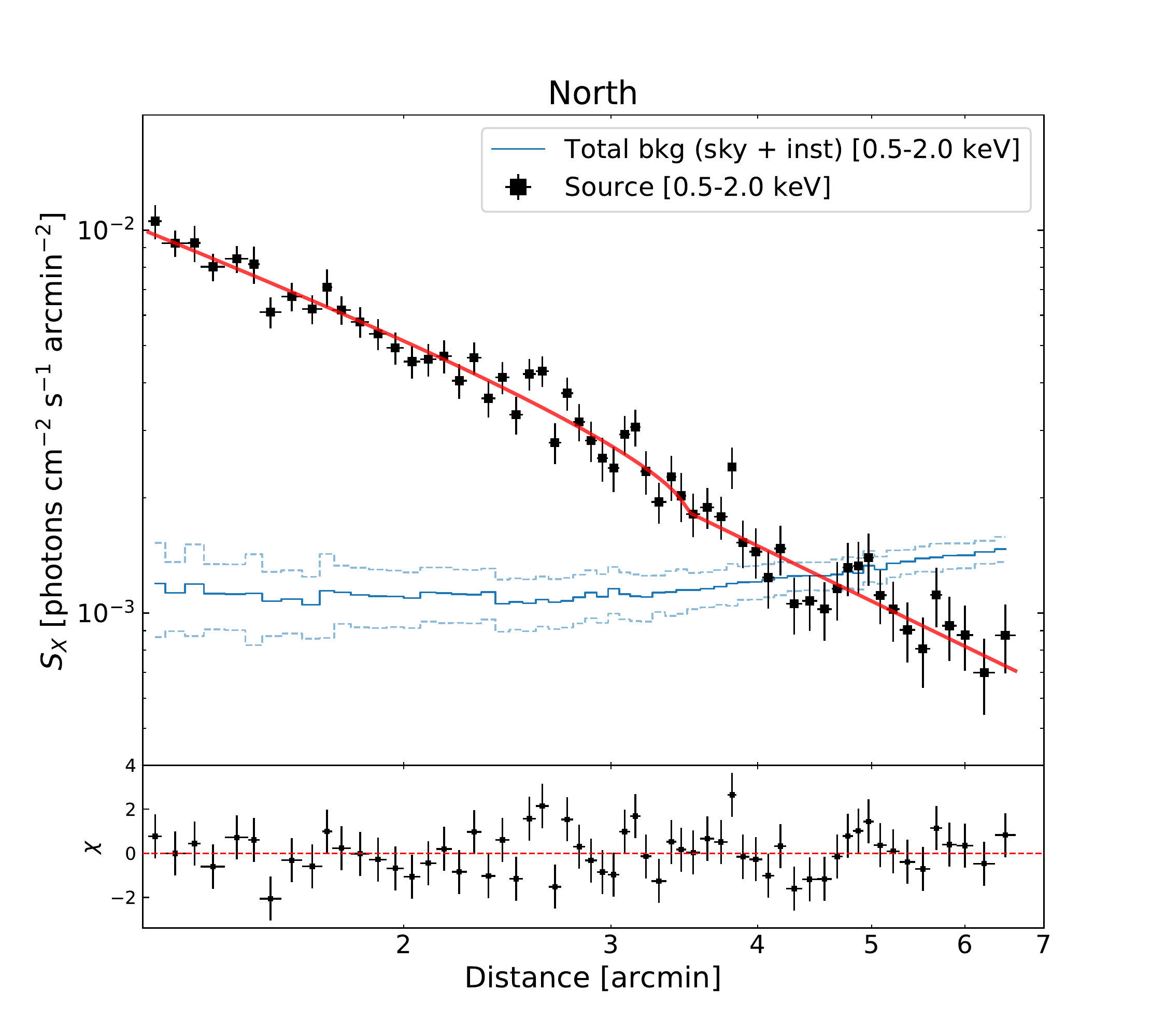}}\\
\caption{Surface brightness profiles across the northern sector. Left: we present the surface brightness across A3411--12 bullet (cold front). Right: we present the surface brightness further north where a hint of a bow shock is detected. The total background level (i.e. instrumental and astrophysical) is shown by the blue line, with the $\pm1\sigma$ uncertainties (blue dashed lines). On the bottom of each panel, the residuals (i.e. $\frac{S_{\rm X,obs}-S_{\rm X,mod}}{\Delta S_{\rm X,obs}}$) are displayed.}\label{fig:sb_northern}
\end{figure*}

\begin{figure*}
\centering
{\includegraphics[width=1.00\textwidth]{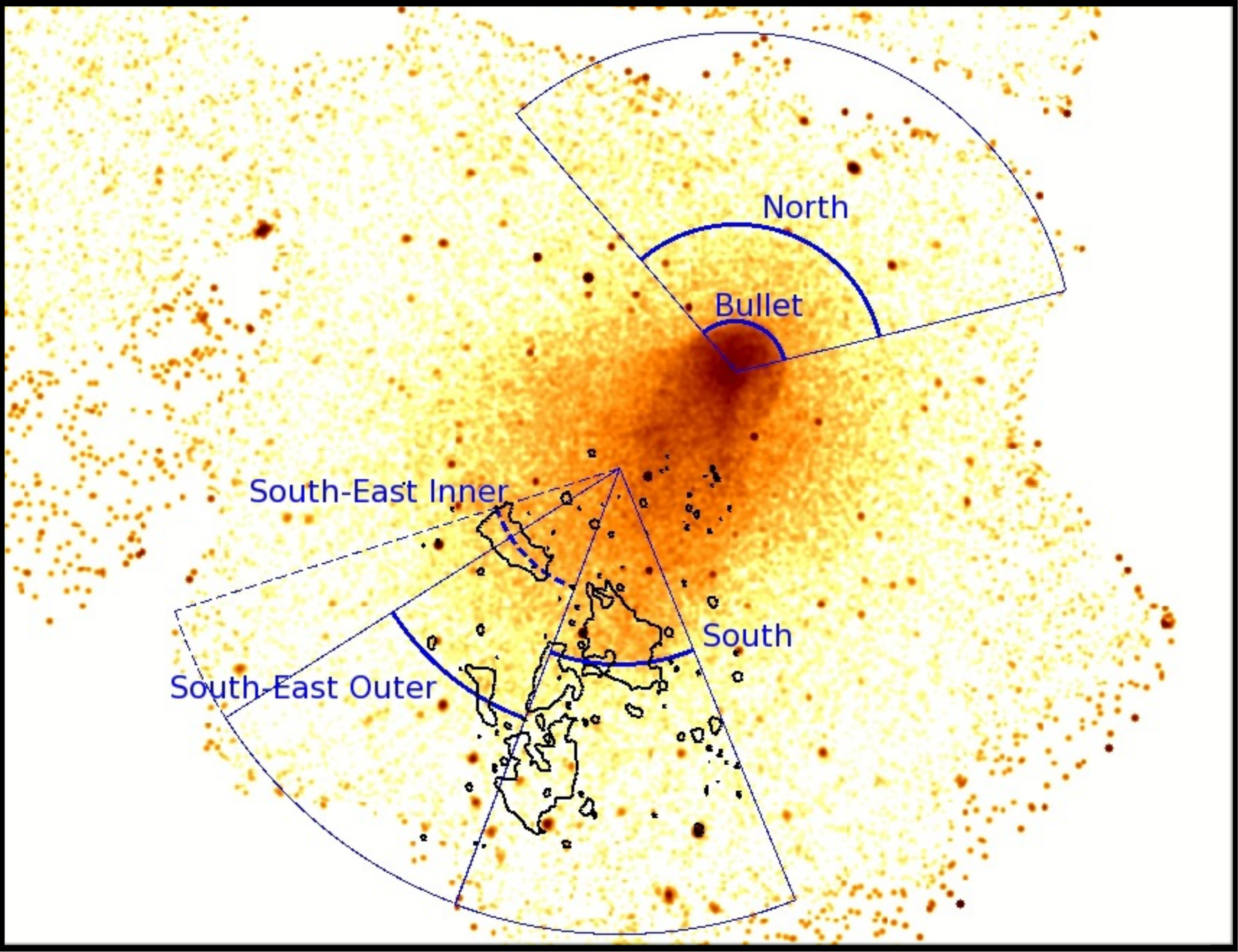}}
\caption{Blue: sectors used for surface brightness extraction.
Black: radio contours indicating the location of radio relics. Black contours display the radio emission of A3411--12 at the $3\sigma_{\rm rms}$ level, with $\sigma_{\rm rms}=0.4~\mu$Jy beam$^{-1}$.}\label{fig:regions_sb}
\end{figure*}

\subsection{Temperature, Pressure and Entropy Maps}

\begin{figure*}
\centering
{\includegraphics[width=0.49\textwidth]{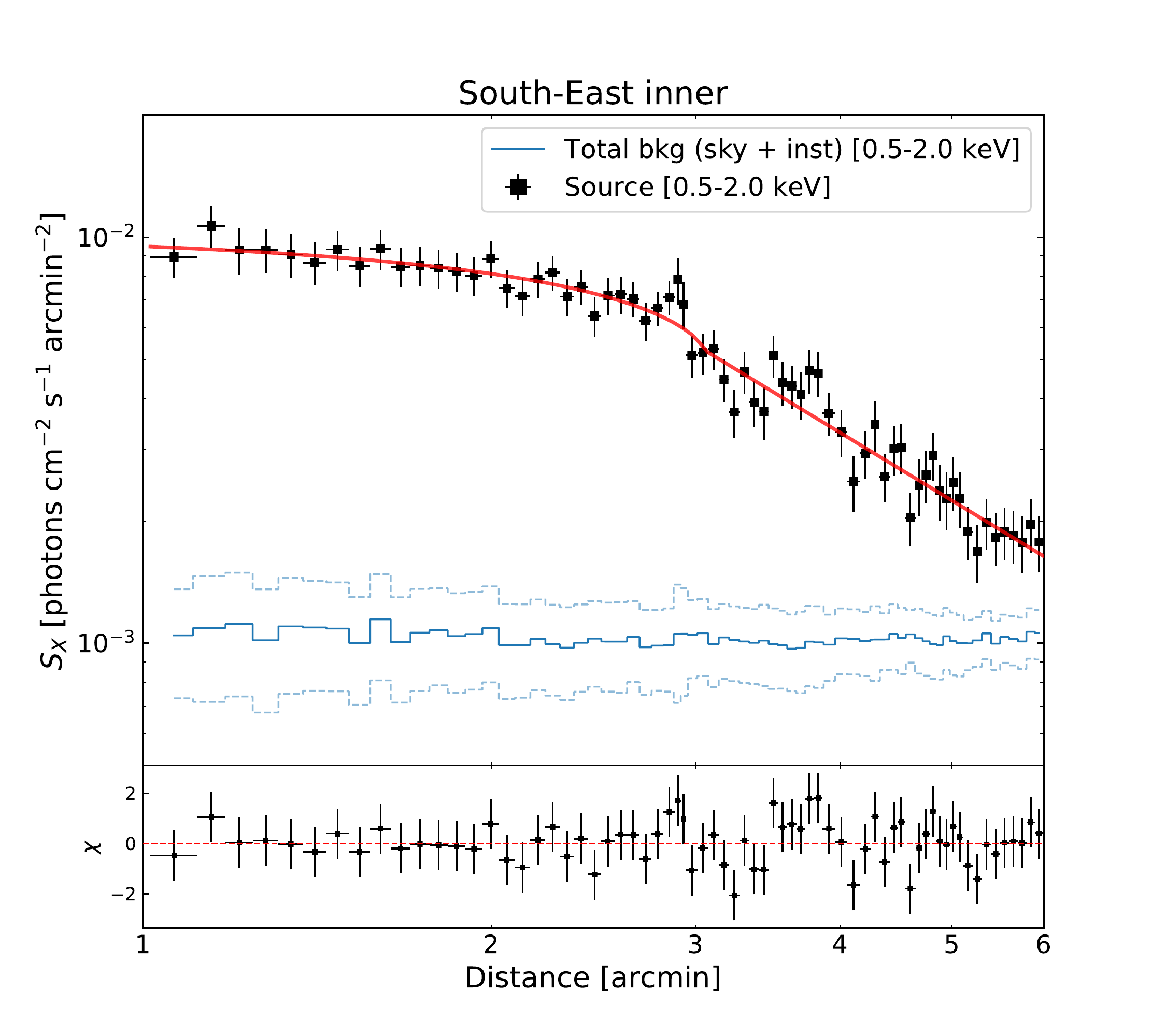}}
\hspace{0.16cm}
{\includegraphics[width=0.49\textwidth]{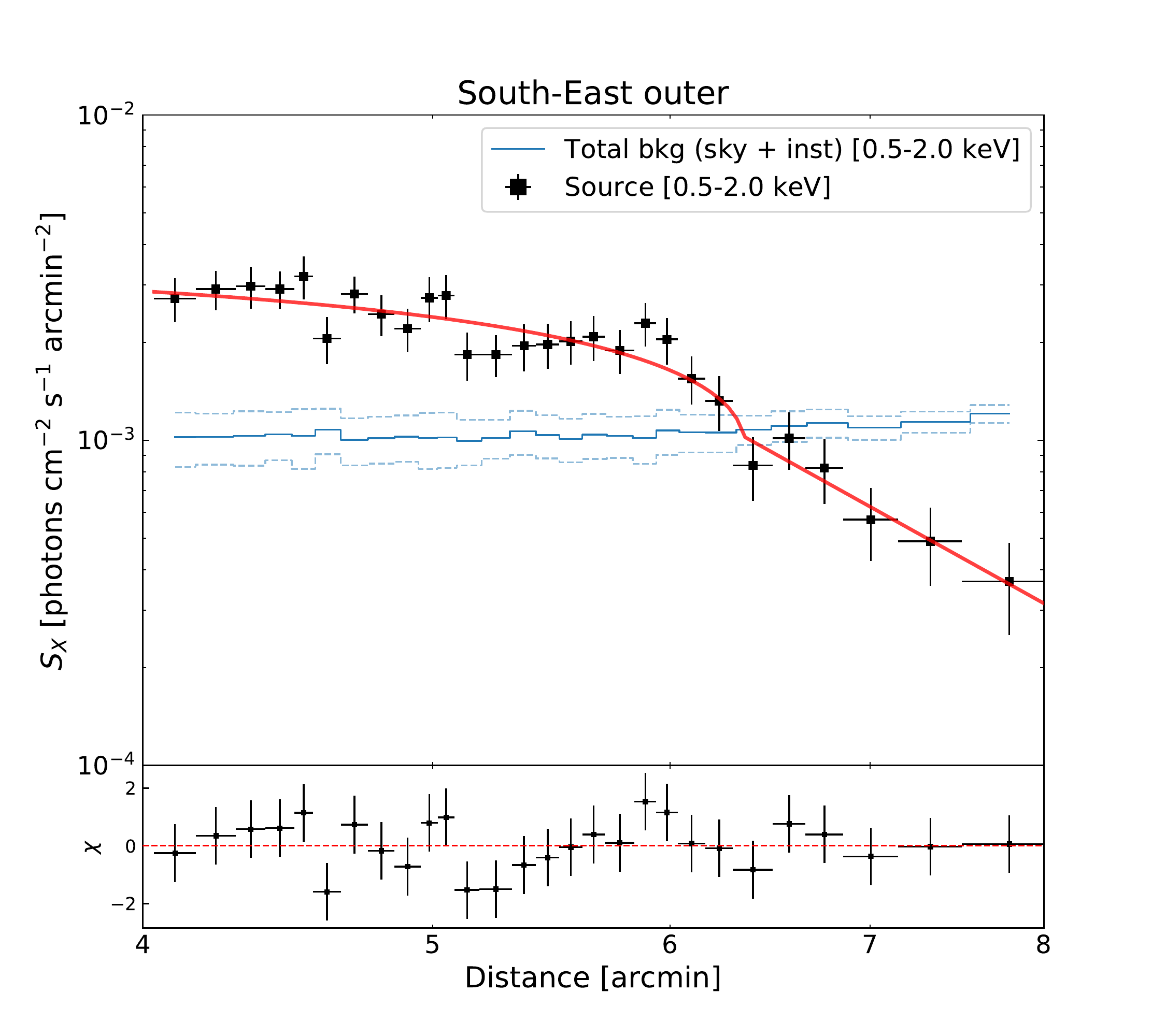}}\\
\caption{Surface brightness profiles across the southeast sector. Left: we present the surface brightness across A3411--12 southeast. Right: we present the surface brightness further south where a hint of an edge is detected. The total background level (i.e. instrumental and astrophysical) is shown by the blue line, with the $\pm1\sigma$ uncertainties (blue dashed lines). On the bottom of each panel, the residuals (i.e. $\frac{S_{\rm X,obs}-S_{\rm X,mod}}{\Delta S_{\rm X,obs}}$) are displayed.}\label{fig:sb_south_east}
\end{figure*}

\begin{figure}
\centering
{\includegraphics[width=0.49\textwidth]{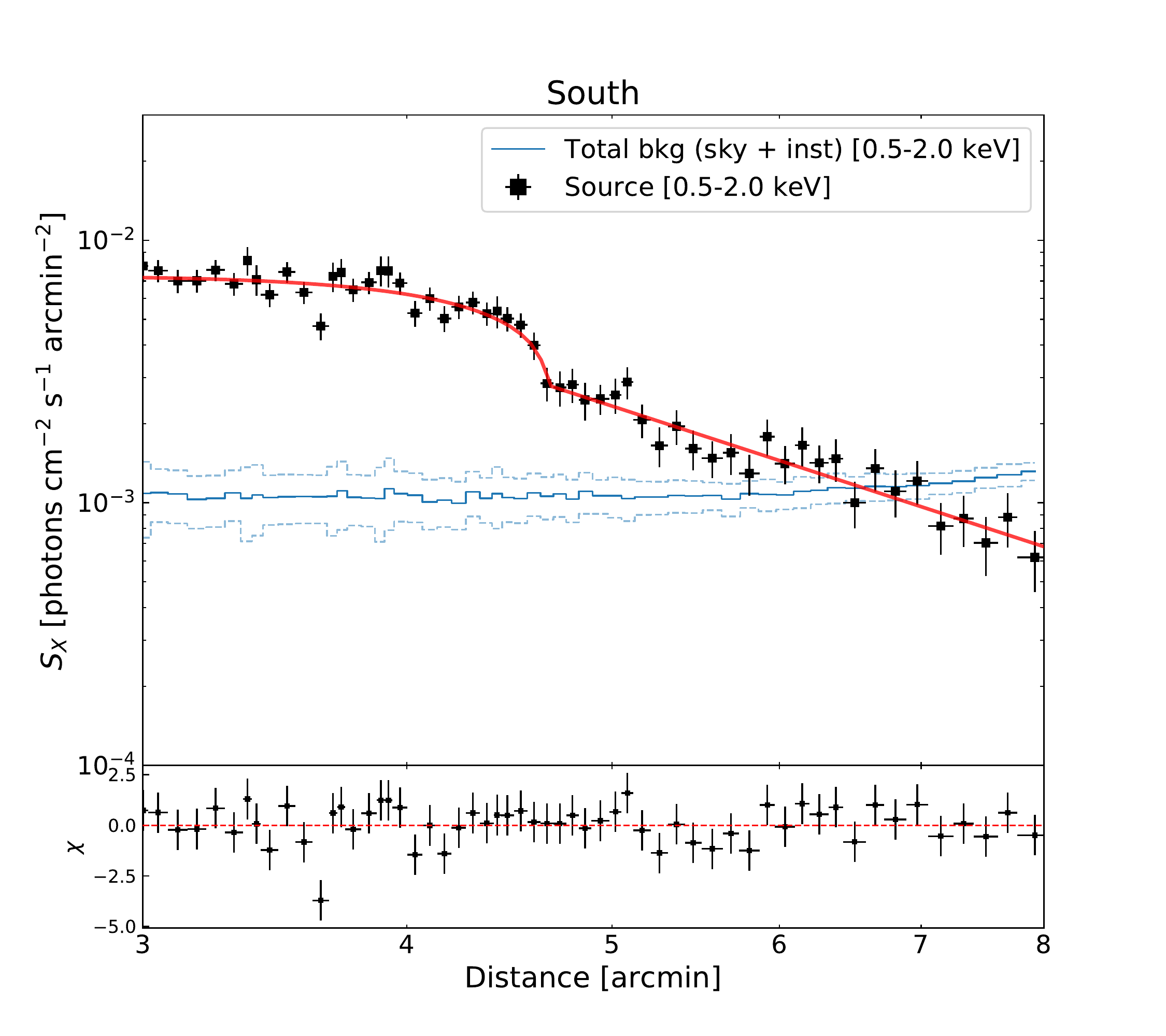}}
\caption{Surface brightness profiles across the south sector. The total background level (i.e. instrumental and astrophysical) is shown by the blue line, with the $\pm1\sigma$ uncertainties (blue dashed lines). On the bottom of each panel, the residuals (i.e. $\frac{S_{\rm X,obs}-S_{\rm X,mod}}{\Delta S_{\rm X,obs}}$) are displayed.}\label{fig:sb_south}
\end{figure}

In this section we present pseudo-density, projected temperature, pseudo-pressure, and pseudo-entropy maps for A3411--12,
extracted from the {\em Chandra} observations. 
For typical cluster
temperatures (k$T$ = 3 -- 10 keV) and metal abundances ($Z$ = 0.1 -- 0.5 $Z_\odot$), the broadband response of 
{\em Chandra} to optically thin thermal
emission from hot gas can be reasonably assumed to be constant with
gas temperature. As an example, for a fixed emission measure, the
0.5--2.5 keV count rate of the {\em Chandra} ACIS-I declines by only $\sim
17\%$ when $kT$ increases from 4 to 12 keV. Therefore, we can ignore
the {\em Chandra} response and assume that the count rate per unit
volume of the gas is directly proportional to the square of the gas
density. Thus, from the surface brightness we can map the projected density of
the cluster, and combining that with a temperature map we can also compute the
pseudo pressure and entropy maps using the following relations:
\begin{eqnarray}
&&n_{\rm e} \propto S^{1/2}, \\
&&P = n_{\rm e} kT \propto S^{1/2} T, \\
&&K = n_{\rm e}^{-2/3} kT \propto S^{-1/3} T, 
\end{eqnarray}
where $S$ and $T$ are the surface brightness and temperature maps.

We extracted spectra in regions that were created using
{\em contbin}, an algorithm for binning X-ray data using
contours on an adaptively smoothed map. 
The generated bins closely follow the surface brightness, 
and are ideal where the surface brightness distribution is not smooth, 
or the spectral properties are expected to follow the surface brightness
\citep{2006Sanders}. The regions were 
selected to have a minimum S/N of 50 in the 0.5--7.0 keV band. Background (sky + detector +
readout) and exposure maps were used. The temperature map was created
by fitting, in the 0.5--7.0 keV band, an absorbed plasma model (XSPEC -- {\em wabs*apec}) to the spectrum data in each
region. NH was fixed to Galactic value of $5.92 \times 10^{20}$ (see Section \ref{section:emmprof}).

In Figure \ref{fig:rho_kT_P_K} we present the projected density,
temperature, pressure, and entropy maps for A3411. The density and
pressure maps present very smooth spatial variations, however the temperature map
presents large variations within relatively small distances. The homogeneity of the pressure map
across the north surface brightness discontinuity shows that the
pressure varies smoothly indicating a cold front, in contrast with
shock fronts which present pressure jumps. The entropy map is
significantly more homogeneous than the temperature map, especially in
the inner regions, suggesting an isentropic process. An isentropic process
is the equivalent to the thermodynamic process which is reversible and adiabatic,
meaning that no heat is dissipated. This suggests that the merger
pushed the low entropy gas back from the core causing it to spread in the
downstream, however, in a mild way, so heat dissipation did not
happen. On the other hand, the inhomogeneity of the temperature map in
the same region suggests that the gas has been mixed. 

\subsection{Shock and Cold Fronts}

For a detailed description of the physics of shock and cold fronts in
galaxy clusters, please refer to \citet{2007Markevitch,2001Vik}. Here we 
discuss briefly the theory behind such phenomena, which will be
relevant in the following analysis.

Let us consider a dense and cold gas cloud moving across a hotter
gas. In Figure \ref{fig:geometry_of_flow}, we show an example of this setup. 
Far upstream from the dense cloud, the gas will be moving
(relative to the dense gas cloud) freely. This region is referred to
as the free stream and will be labeled with the index 1 (See Figure \ref{fig:geometry_of_flow}). The hot gas decelerates as it approaches the dense gas cloud, approaching zero
velocity at the edge of the dense cloud. This region is referred to
as the stagnation point and will be labeled with the index 0. 
The density discontinuities in cold fronts 
form whenever a gas density peak encounters a flow of ambient gas, causing a contact
discontinuity to quickly form. Furthermore, if the
velocity of the dense gas cloud exceeds the sound speed of the hot
gas, a bow shock forms at some distance upstream from the dense
cloud. The region just inside the bow shock will be indexed as 2 
(see Fig. \ref{fig:geometry_of_flow} for a visual description of the regions
discussed above).
The ratio of pressures in the free stream (1) and at the stagnation
point (0) is a function of the cloud speed $v$ \citep[Section 114 of][]{1959Landau}:

\begin{eqnarray}
\frac{p_0}{p_1}=\left\{ \begin{array}{ll}
\renewcommand{\arraystretch}{3}
\left( 1 + \frac{\gamma - 1}{2}M_1^2
\right)^\frac{\gamma}{\gamma-1},	
			& \mbox{\hspace{0.1cm} $M_1 \leq 1$}\\
\left(\frac{\gamma +
  1}{2}\right)^\frac{\gamma+1}{\gamma-1} M_1^2\left(\gamma -
\frac{\gamma-1}{2M_1^2}\right)^\frac{-1}{\gamma-1},	
			& \mbox{\hspace{0.1cm} $M_1 > 1$}\\
\end{array}
\right.
\label{eq:p0_p1}
\end{eqnarray}
where $M_1 = v/c_1$ is the Mach number in the free stream and $\gamma$
is the adiabatic index of the gas. The pressure
ratio dependence on the square of the Mach number leads to a large
increment in the pressure ratio for relatively small changes in the
velocity of the cloud, therefore the cloud velocity can be measured
rather accurately even if the pressure uncertainties are relatively high. 

\begin{deluxetable}{lccc}[!t]
\tablecaption{Density jumps and Mach numbers} 
\tablewidth{0pt} 
\tablehead{ 
\colhead{Sector} &
\colhead{$r_{\rm shock}$ (arcmin)} & 
\colhead{$C$} &
\colhead{$M$}
}
\startdata 
Bullet & $1.20^{+0.01}_{-0.03}$ & $1.99_{-0.05}^{+0.09}$ & cold-front \\
Northern & $3.48^{+0.61}_{-0.71}$ & $1.22_{-0.14}^{+0.20}$ & $1.15_{-0.09}^{+0.14}$ \\
Southern & $4.65^{+0.04}_{-0.04}$ & $1.70_{-0.09}^{+0.09}$ & cold-front \\
Southeast inner & $3.04^{+0.15}_{-0.18}$ & $1.19_{-0.13}^{+0.21}$ & $1.13_{-0.08}^{+0.14}$ \\
Southeast outer & $6.34^{+0.40}_{-0.26}$ & $1.31_{-0.17}^{+0.22}$ & $1.21_{-0.12}^{+0.15}$
\enddata
\tablecomments{Columns list best fit values for the parameters
given by Equations \ref{eq:compression} and \ref{eq:compression_RH}.} 
\label{tab:C_M}
\end{deluxetable}

\begin{deluxetable}{lccc}[!t]
\tablecaption{Temperature} 
\tablewidth{0pt} 
\tablehead{ 
\colhead{Sector} &
\colhead{$r_{\rm in}$ (arcmin)} & 
\colhead{$r_{\rm out}$ (arcmin)} &
\colhead{k$T$ (keV)}
}
\startdata 
N Cold-front & 0 & 1.20 & $5.15_{-0.18}^{+0.19}$ \\
N Downstream & 1.20 & 3.48 & $7.40_{-0.62}^{+0.62}$ \\
N Upstream & 3.48 & 8 & $5.06_{-0.60}^{+0.74}$ \\
S Cold-front & 0 & 4.65 & $6.50_{-0.32}^{+0.33}$ \\ 
S Downstream & 4.65 & 11 & $6.40_{-0.96}^{+1.32}$ \\ 
SE I Downstream & 0 & 3.04 & $6.26_{-0.43}^{+0.44}$ \\
SE I Upstream & 3.04 & 11 & $4.56_{-0.39}^{+0.43}$ \\
SE O Downstream & 0 & 6.34 & $5.81_{-0.39}^{+0.40}$  \\
SE O Upstream & 6.34 & 11 & $2.54_{-0.96}^{+1.91}$ 
\enddata
\tablecomments{Columns list the sector used to extract the temperature (N -- north, S -- south,
SE I -- southeast inner, SE O -- southeast outer), the inner and outer radii, and temperature.}
\label{tab:temp_sectors}
\end{deluxetable}

As mentioned earlier, if the speed of a blunt body exceeds the speed of sound, a bow shock
forms at some distance upstream. The shape of this structure is
consistent with an ellipse centered on the center of curvature of the
cold front. If the surface brightness discontinuity is interpreted as
a shock front, it is straightforward to derive the expected
temperature jump, the shock propagation velocity and the velocity of
the gas behind the shock, using the Rankine-Hugoniot shock equations
\citep[Section 85 of][]{1959Landau}:
\begin{eqnarray}
\frac{\rho_2}{\rho_1} &=& \frac{(1+\gamma)M_1^2}{2+(\gamma -
  1)M_1^2}, \label{eq:Rankine-Hugoniot_1} \\
\frac{T_2}{T_1} &=& \frac{2\gamma M_1^2 - \gamma + 1}{\gamma + 1}\frac{\rho_1}{\rho_2}.
\label{eq:Rankine-Hugoniot_2}
\end{eqnarray}
where $\rho_2/\rho_1$ and $T_2/T_1$  are the ratios of densities and temperatures in the downstream (2) and free stream (1) regions, respectively. 

\subsubsection{Modeling the density jumps}

Following \citet{2009Owers}, we can fit the surface brightness profile across a shock
assuming spherical symmetry for the gas density profile, which is
given by two power laws (broken power law):
\begin{eqnarray}
n(r)=\left\{ \begin{array}{ll}
\renewcommand{\arraystretch}{3}
C n_{0,2} \left(\frac{r}{r_{\rm
    shock}}\right)^{-\alpha_1},	
			& \mbox{\hspace{0.9cm} $r \leq r_{\rm
  shock}$}\\
n_{0,2} \left(\frac{r}{r_{\rm shock}}\right)^{-\alpha_2},	
			& \mbox{\hspace{0.9cm} $r > r_{\rm shock}$}\\
\end{array}
\right.
\label{eq:compression}
\end{eqnarray}
where $C$ is the density compression ($\rho_2/\rho_1$), and $r_{\rm shock}$ 
is the radius at the shock (where the surface brightness
discontinuity is located). $C$ is directly related to the
Mach number via the Rankine-Hugoniot equations presented in Equation
(\ref{eq:Rankine-Hugoniot_1}). For mono-atomic gas, $\gamma = 5/3$, and:
\begin{eqnarray}
C = \frac{4 M_1^2}{3 + M_1^2}.
\label{eq:compression_RH}
\end{eqnarray}

The gas density at the shock upstream region is typically very low,
which makes measuring the temperature jump quite
difficult, A3411--12 being no exception despite our very good {\em Chandra} data.

\subsubsection{Northern Cold and Shock Fronts}

Cold fronts are found in many galaxy clusters \citep[Bullet Cluster,
A2029, A2204, RXJ1720, Ophiuchus, A2142, A3667, A1644, A520, and many
more][]{2007Markevitch}. While cold fronts may be the result of many different
physical events in the cluster, the density discontinuities 
in them form for the same basic reason: whenever 
a gas density peak encounters a flow of ambient gas, a contact
discontinuity quickly forms.

Here, we measure the surface brightness profile in a wedge towards the
north of A3411, centered on the bright northern cool core. The left panel of Figure
\ref{fig:sb_northern} shows the cold front signature, as
modeled by a broken power law (Equation \ref{eq:compression}). The density jump factor 
is $C = 1.99$. The right panel of Figure \ref{fig:sb_northern} shows what may be
the bow shock, at 2$'$.3 upstream from the cold front, modeled as another broken power law. 
The density jump is $C = 1.22_{-0.14}^{+0.20}$, which implies $M = 1.15_{-0.09}^{+0.14}$ (with a 90\% confidence upper limit of $M<1.6$)
if we interpret this density discontinuity as a shock front. 
We used a Bayesian information criterion (BIC) to compare a single power
law model to a broken power law model. For the single power law model,
we obtain BIC = 75.9 ($\chi^2$ = 67.75). For the broken power law model,
we obtain BIC = 75.2 ($\chi^2$ = 54.74). Despite the slightly lower BIC in favor of the density jump, no clear conclusion can be drawn from the current data.
The density jumps and Mach numbers for all sectors
are presented in Table \ref{tab:C_M}.

Table \ref{tab:temp_sectors} shows the temperature at the regions presented in Figure \ref{fig:regions_sb}, which are determined by the surface brightness jumps. Towards the north, the cool core becomes very clear in the surface brightness profile, as well as the temperature jump from inside to outside of the stagnation
point. Further north, the temperature jump and density discontinuities are only suggestive of a shock front.
Indeed, computing the Mach number using temperature (Equation \ref{eq:Rankine-Hugoniot_2}) associated with all three suggestive shock fronts (in the Northern, Southeast Inner, and Outer sectors) leads to unconstrained results.   

\subsubsection{Southeast Radio Relic, Cold Front, and Possible Shock Front}

Measuring the surface brightness towards the 
southeast of A3411--12, we also 
see the suggestion of an upstream
bow shock (right panel of Figure \ref{fig:sb_south_east}).

The surface brightness discontinuity presented in the southeast inner region (see left panel of Figure \ref{fig:sb_south_east}) has been associated with a shock-front, responsible for electron re-acceleration producing the bright radio relic \citep[][see also Section~\ref{sec:formationofrelics}]{2017vanWeerenNatAst}. 
For this sector we measure $C = 1.19_{-0.13}^{+0.21}$ and $M = 1.13_{-0.08}^{+0.14}$ (with a 90\% confidence upper limit of $M<1.6$).

\subsubsection{Southern Cold Front}

Measuring the surface brightness towards the bright X-ray clump in the
south of A3411--12, we also see another cold front signature, as the
X-ray surface brightness profile
is very well modeled by a broken power law (Figure
\ref{fig:sb_south}). Measuring an upstream
bow shock, however, is not possible due to the very low statistic at
such a large distance from the cluster X-ray bright regions. 
For this sector we measure $C = 1.70_{-0.09}^{+0.09}$.

%%%%%%%%%%%%%%%%%%%%%%%%%%%%%%%%%%%%%%%%%%%%%%%%%%%%%%%%%%%%%%%%%%%%%%%%%%%%%%%%%%
%%%%%%%%%%%%%%%%%%%%%%%%%%%%%%%%%%%%%%%%%%%%%%%%%%%%%%%%%%%%%%%%%%%%%%%%%%%%%%%%%%
%%
%%                            MERGING SCENARIO
%%
%%%%%%%%%%%%%%%%%%%%%%%%%%%%%%%%%%%%%%%%%%%%%%%%%%%%%%%%%%%%%%%%%%%%%%%%%%%%%%%%%%
%%%%%%%%%%%%%%%%%%%%%%%%%%%%%%%%%%%%%%%%%%%%%%%%%%%%%%%%%%%%%%%%%%%%%%%%%%%%%%%%%%

\section{Merging Scenario}

Optical, X-ray, and radio data indicate that A3411--12 is undergoing a major
merger. \citet{2017vanWeerenNatAst} showed that
the optical data indicates a clear bimodal galaxy
distribution, with velocity dispersion indicating a merger with a 1:1
mass ratio, and a
radial velocity difference between the two peaks compatible with zero,
suggesting a merger on the plane of the sky. The X-ray data show a
bullet-like cool core in the north with extended diffuse X-ray
emission in the south, also highly suggestive of a merger happening
mostly in the direction south-north on the plane of the sky. Cold
fronts in the south and north regions also support this, as well as
what seems be a bow shock upstream from the north cold front
density discontinuity. Assuming this density discontinuity to be a
shock, we measure a Mach number of $M = 1.15_{-0.09}^{+0.14}$. 
The radio data show a large, Mpc scale relic in the south, a typical
signature found in the outskirts of many merging clusters.

Thanks to the very good {\em Chandra} X-ray data, we are able identify and
constrain what seems to be a shock in the south (Southeast Outer sector -- it is important to note that we cannot constrain the temperature, therefore the nature of the density jump), with a Mach number of $M = 1.21_{-0.12}^{+0.15}$. 
In \citet{2017vanWeerenNatAst}, the best fit for the density jump at the location  of the radio relic gives $M = 1.2$, 
with a 90\% confidence upper limit of $M < 1.4$, also suggesting a low Mach number. 
Here we measure at the location  of the radio relic presented in \citet{2017vanWeerenNatAst} (Southeast Inner sector) $M = 1.13_{-0.08}^{+0.14}$.

\subsection{Merger Analogs}

To model the dynamics of the merger we used the method of
\citet{WittmanAnalogs2019}, who uses the projected separation, relative
line-of-sight velocity, and masses to select analog systems from a
cosmological N-body simulation.  We followed that work in using the Big
Multidark Planck (BigMDPL) Simulation \citep{BigMDPL2016} hosted on
the \textit{cosmosim.org} website, but we updated the cluster
parameters as follows. First, the X-ray morphology strongly suggests
that the subclusters are still outgoing, so we eliminate analog
systems that are in the returning phase. Second, our $M_{500}$
estimate implies a total virial mass $\approx 1.4 \times M_{500} =
10\times10^{14}$ M$_\odot$.  This is lower than the $16\times10^{14}$
M$_\odot$ total mass used by \citet{WittmanAnalogs2019}, who noted
that the only available mass estimate available then was a dynamical
mass likely to be biased high.  Because the velocity dispersions of
the two subclusters are almost equal and our X-ray data do not
constrain the mass ratio, we search for analog systems with subcluster
virial masses of $(5\pm2.5)\times10^{14}$ M$_\odot$.

The resulting constraints are: the subcluster separation vector is
$>74$ ($>60$) degrees from the line of sight at 68\% (95\%) confidence;
the time since pericenter passage is 460-790 (340-820) Myr at the same
confidence levels; the maximum relative speed reached near pericenter passage,
$v_{\rm max}$, is 2000--2500 (1900--2800) $\rm km~s^{-1}$; and the relative speed
at the time of observation is 540--1100 (320--1400) $\rm km~s^{-1}$.  Note that
the maximum relative speed of the analog halos is likely to be
underestimated due to confusion in assigning particles to overlapping
halos \citep{WittmanAnalogs2019}.

These estimates are consistent with the shock position as
follows. Hydrodynamical simulations of a merging cluster
\citep{2007Springel} indicate that shocks are launched from near the
center of mass (CM) around the time of pericenter passage. The maximum speed
$v_{\rm max}$ sets the speed of the shock front; over time the 
subclusters slow substantially (and eventually fall back) while the
shock slows little. Hence the analogs predict that shock fronts have
been traveling at $\gtrsim 1200$ $\rm km~s^{-1}$ in CM coordinates for 650 Myr,
for a distance of $\gtrsim 830$ kpc.  Because the analogs also predict
that the separation vector is close to the plane of the sky, we expect
the projected separation between shock and CM to be $\gtrsim$ 800 kpc.
In fact we find $\sim$ 1.1 Mpc, which indeed
requires a projected separation not much less than the physical
separation, {\it as well as} the analog speed being biased low by
several hundred $\rm km~s^{-1}$. 

If the shock propagates at $\approx 1600$ $\rm km~s^{-1}$ in CM coordinates as
suggested by the 1.1 Mpc separation, this implies $\approx 3000$ $\rm km~s^{-1}$
relative to the unshocked gas, or $\mathcal{M}\approx 3$\footnote{For k$T \sim 5$ keV (the measured temperature in the north upstream region) the sound speed is  $\sim 1000 \rm ~km~s^{-1}$.}. The smaller
Mach number we find could be due to line-of-sight projections of
different parts of the 3-D shock front. It could also be due to
slowing of the shock over time: although \citet{2007Springel} found
that the slowing was only about 10\%, their simulations extended only
300 Myr past pericenter, while we are observing A3411 much later,
$\approx 650$ Myr after pericenter passage.  Addressing these issues will
require detailed hydrodynamical simulations, beyond the scope of this
paper.

%%%%%%%%%%%%%%%%%%%%%%%%%%%%%%%%%%%%%%%%%%%%%%%%%%%%%%%%%%%%%%%%%%%%%%%%%%%%%%%%%%
%%%%%%%%%%%%%%%%%%%%%%%%%%%%%%%%%%%%%%%%%%%%%%%%%%%%%%%%%%%%%%%%%%%%%%%%%%%%%%%%%%
%%
%%                               RELICS
%%
%%%%%%%%%%%%%%%%%%%%%%%%%%%%%%%%%%%%%%%%%%%%%%%%%%%%%%%%%%%%%%%%%%%%%%%%%%%%%%%%%%
%%%%%%%%%%%%%%%%%%%%%%%%%%%%%%%%%%%%%%%%%%%%%%%%%%%%%%%%%%%%%%%%%%%%%%%%%%%%%%%%%%

\section{Formation of Radio Relics}
\label{sec:formationofrelics}
From our analysis of the density jumps across surface brightness discontinuities we
conclude that if they are indeed shocks, they are very mild. However, radio relics extend 
over Mpcs in the A3411--12 system (see radio emission (red) in Figure 
\ref{fig:optical_xray_radio}). The fact that we observe very extended radio relics
in a cluster with such low-Mach number shocks is indicative that a population of 
energetic electrons already existed over extended regions of the cluster.

The southeast inner edge has been discussed in \cite{2017vanWeerenNatAst} where it is argued that this edge is a shock front where particles from a nearby tailed radio galaxy are being re-accelerated. While there is evidence for a mild density jump at this location, the {\em Chandra} data are not deep enough to confirm the presence of a temperature jump. This means that in principle this edge could also trace a cold front. Given that this edge traces a relic, a shock interpretation is more likely, as this has been confirmed for numerous other relics \citep[e.g.,][]{2019vanWeeren}. However, a cold front (contact discontinuity) might also be able to explain some of the observed radio features. %The measured upper limits on the width of these cold front are often smaller than the mean free paths of particles, which suggests that thermal conductivity and particle diffusion are suppressed.
Magnetic field lines are thought to be stretched along the cold front interface \citep{2006Lyutikov,2008Dursi}. An  alignment of magnetic field might explain the increase of the observed polarization fraction of the radio relic at this location. If the ICM magnetic field is also locally enhanced at the cold front, this will result in a flattening of the radio spectral index, providing that the underlying spectrum contains a spectral cutoff (i.e., is curved). A higher magnetic field strength will then ``illuminate'' a different part of the underlying electron spectrum, causing the observed spectral index to flatten \citep[][]{2012KatzStone}. A curved spectrum with a spectral break is naturally expected due to electron energy losses in the tail of the AGN. Although a shock scenario remains more likely, future temperature measurements will be important to confirm or rule out a cold-front scenario for the origin of the relic.

%%%%%%%%%%%%%%%%%%%%%%%%%%%%%%%%%%%%%%%%%%%%%%%%%%%%%%%%%%%%%%%%%%%%%%%%%%%%%%%%%%
%%%%%%%%%%%%%%%%%%%%%%%%%%%%%%%%%%%%%%%%%%%%%%%%%%%%%%%%%%%%%%%%%%%%%%%%%%%%%%%%%%
%%
%%                               CONCLUSIONS
%%
%%%%%%%%%%%%%%%%%%%%%%%%%%%%%%%%%%%%%%%%%%%%%%%%%%%%%%%%%%%%%%%%%%%%%%%%%%%%%%%%%%
%%%%%%%%%%%%%%%%%%%%%%%%%%%%%%%%%%%%%%%%%%%%%%%%%%%%%%%%%%%%%%%%%%%%%%%%%%%%%%%%%%

\section{Conclusions}

In this paper we presented deep {\em Chandra} observations of the merging cluster
A3411--12. This remarkable cluster hosts the most compelling evidence 
for electron re-acceleration at cluster shocks to date. 
We present gas temperature, X-ray luminosity, gas and total
masses, and gas fraction profiles. We computed the shock strength 
using density jumps to conclude that the Mach number is small ($M \leq 1.15_{-0.09}^{+0.14}$). 
We also presented pseudo-density, projected temperature, pseudo-pressure, and
pseudo-entropy maps. Based on the pseudo-entropy map we
conclude that the cluster is undergoing a mild merger,
consistent with the small Mach number. On the other hand, radio relics span 
over Mpcs in the A3411--12 system, which indicates that a population of 
energetic electrons already existed over extended regions of the cluster. 
In the southeast of the system there is evidence for a mild density jump,
however our {\em Chandra} data are not deep enough to confirm the presence 
of a temperature jump. Therefore, we cannot determine if this edge traces a cold front or a shock. Future higher precision temperature measurements are therefore important to test the shock re-acceleration scenario for radio relic formation.

\acknowledgments

We thank Joseph DePasquale for creating the spectacular image displayed in Figure 1.
F.A.-S. acknowledges support from {\em Chandra} grant G05-16133X. 
R.J.vW. acknowledges support from the VIDI research programme with project number 639.042.729, which is financed by the Netherlands Organisation for Scientific Research (NWO). G.D.G acknowledges support from the ERC Advanced Investigator programme NewClusters 321271.
W.R.F., and C.J. are supported by the Smithsonian Institution.
C.J. acknowledges support from {\em Chandra} grant G016619003.
M.J.J. acknowledges support for the current research from the National Research Foundation of Korea under the programs 2017R1A2B2004644 and 2017R1A4A1015178. D.R. and H.K. acknowledge support for the current research from the National Research Foundation of Korea under the programs 2016R1A5A1013277 and 2017R1A2A1A05071429.
V.M.P. acknowledges partial support for this work from grant PHY 1430152; Physics Frontier Center/ JINA Center for the Evolution of the Elements (JINA-CEE), awarded by the US National Science Foundation (NSF). A.S. acknowledges support from the Clay Fellowship. Part of this was work performed under the auspices of the U.S. DOE by LLNL under Contract DE-AC52-07NA27344.

\clearpage

\appendix
\renewcommand\thefigure{\thesection.\arabic{figure}}

\section{Density jumps and Mach number distributions}\label{apdx:densMach}

Here we present the density jumps and Mach number distributions from Figures  \ref{fig:sb_northern}, \ref{fig:sb_south_east} and \ref{fig:sb_south}. They show the distribution of solutions for the fitted parameters from the X-ray surface brightness profiles across the wedges presented in those figures. 

\setcounter{figure}{0}
\begin{figure*}[!h]
\centerline{
%\vspace{-0.5cm}
\includegraphics[width=0.50\textwidth]{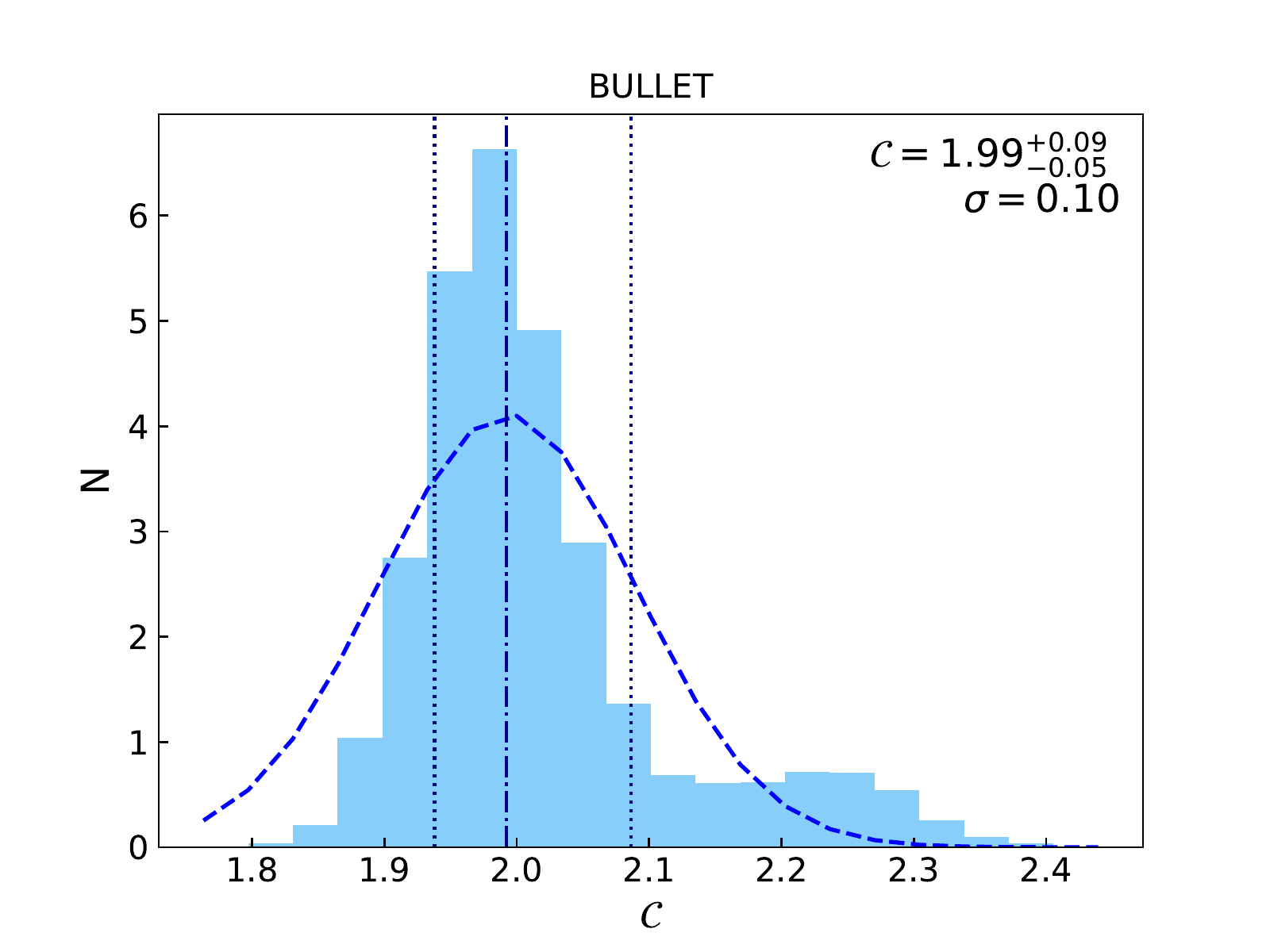}
\includegraphics[width=0.50\textwidth]{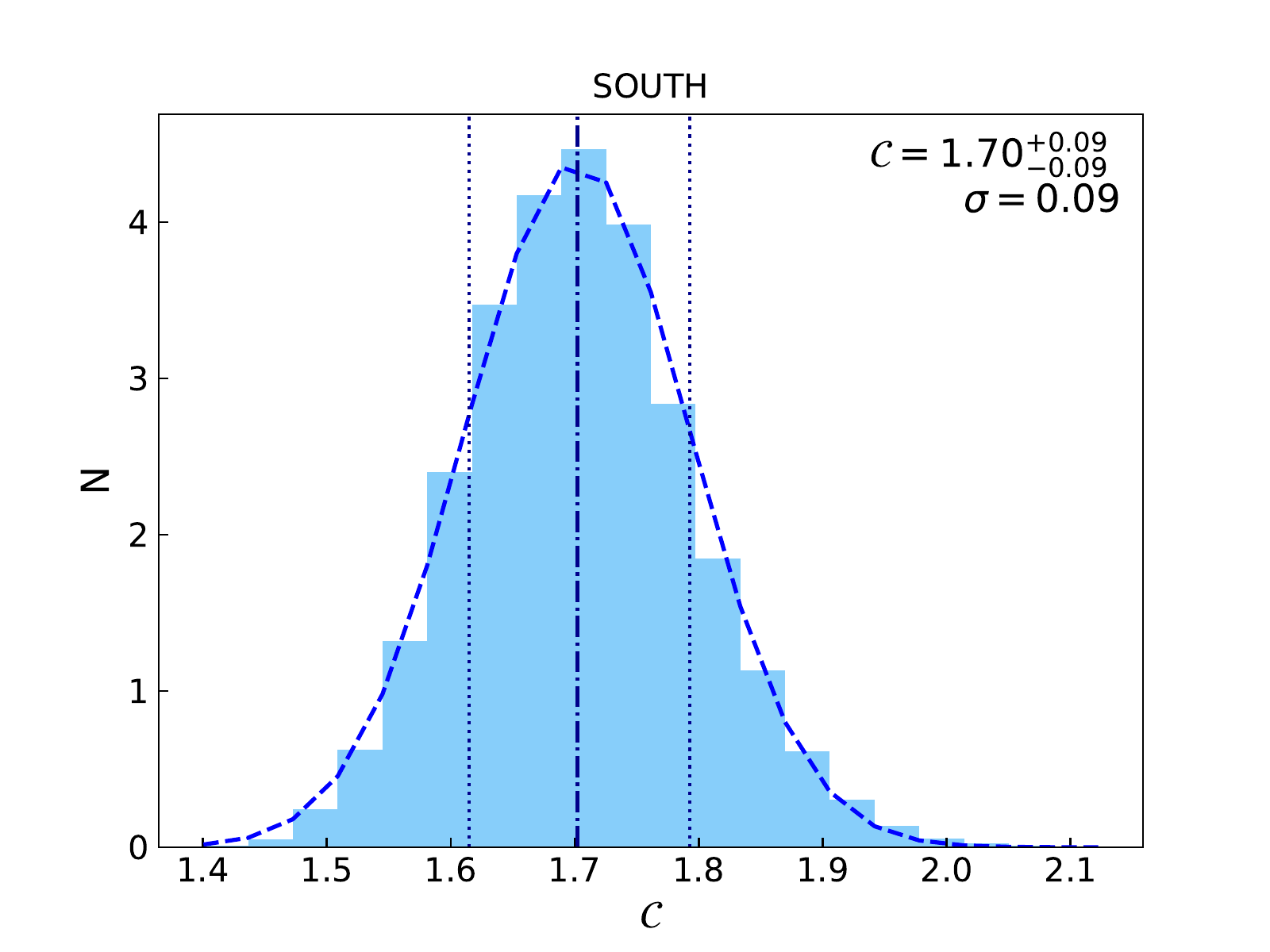}
}
\caption{\small{Distribution of density jumps for the cold fronts in the northern and southern sectors. Left: distribution of density jumps from the MCMC simulations from the cold-front in the northern sector (the bullet). Right: Same as the left panel, except for the southern cold-front.
}}\label{fig:dist_north}
\end{figure*}

\begin{figure*}[!b]
\centerline{
%\vspace{-0.5cm}
\includegraphics[width=0.50\textwidth]{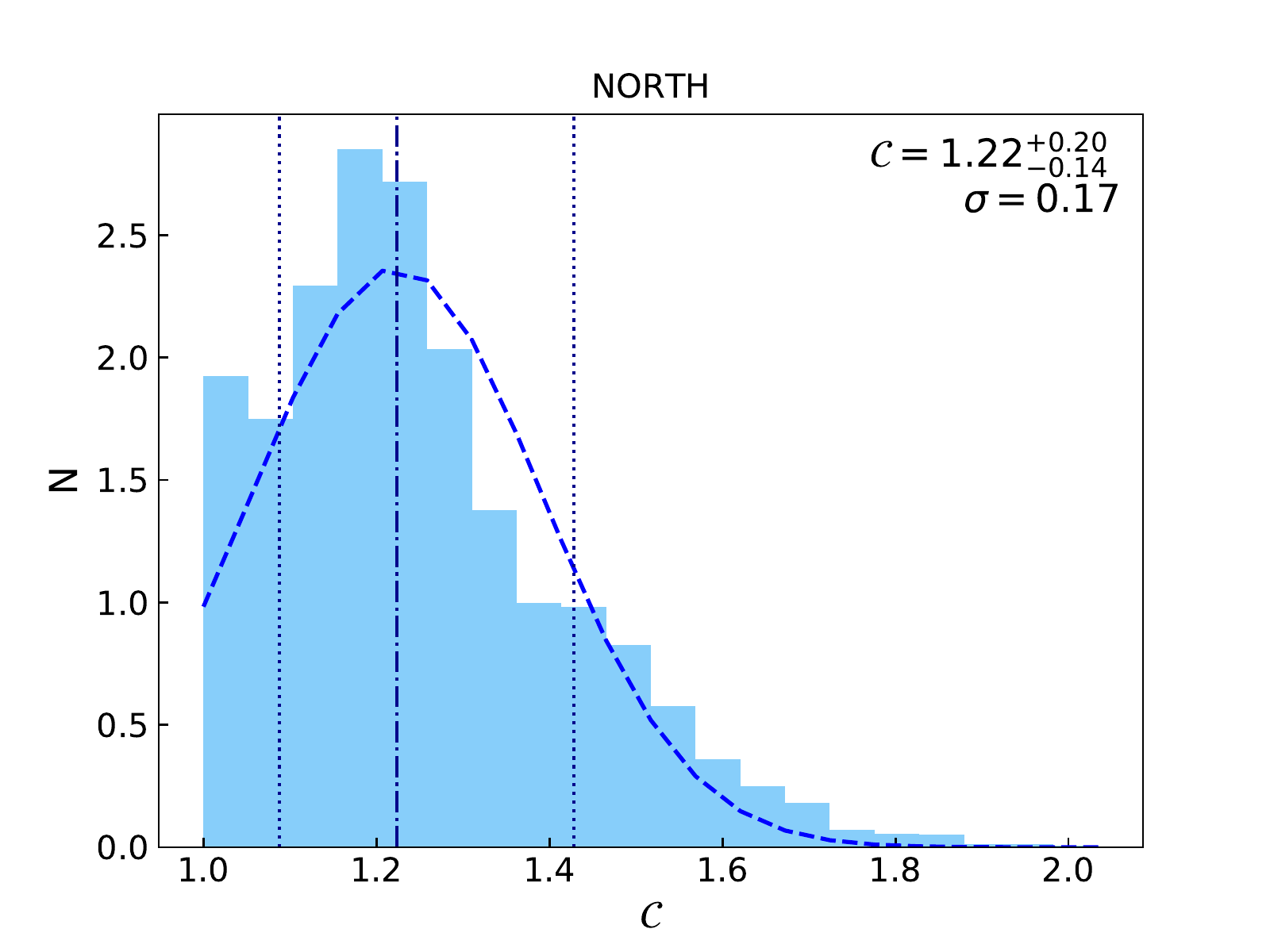}
\includegraphics[width=0.50\textwidth]{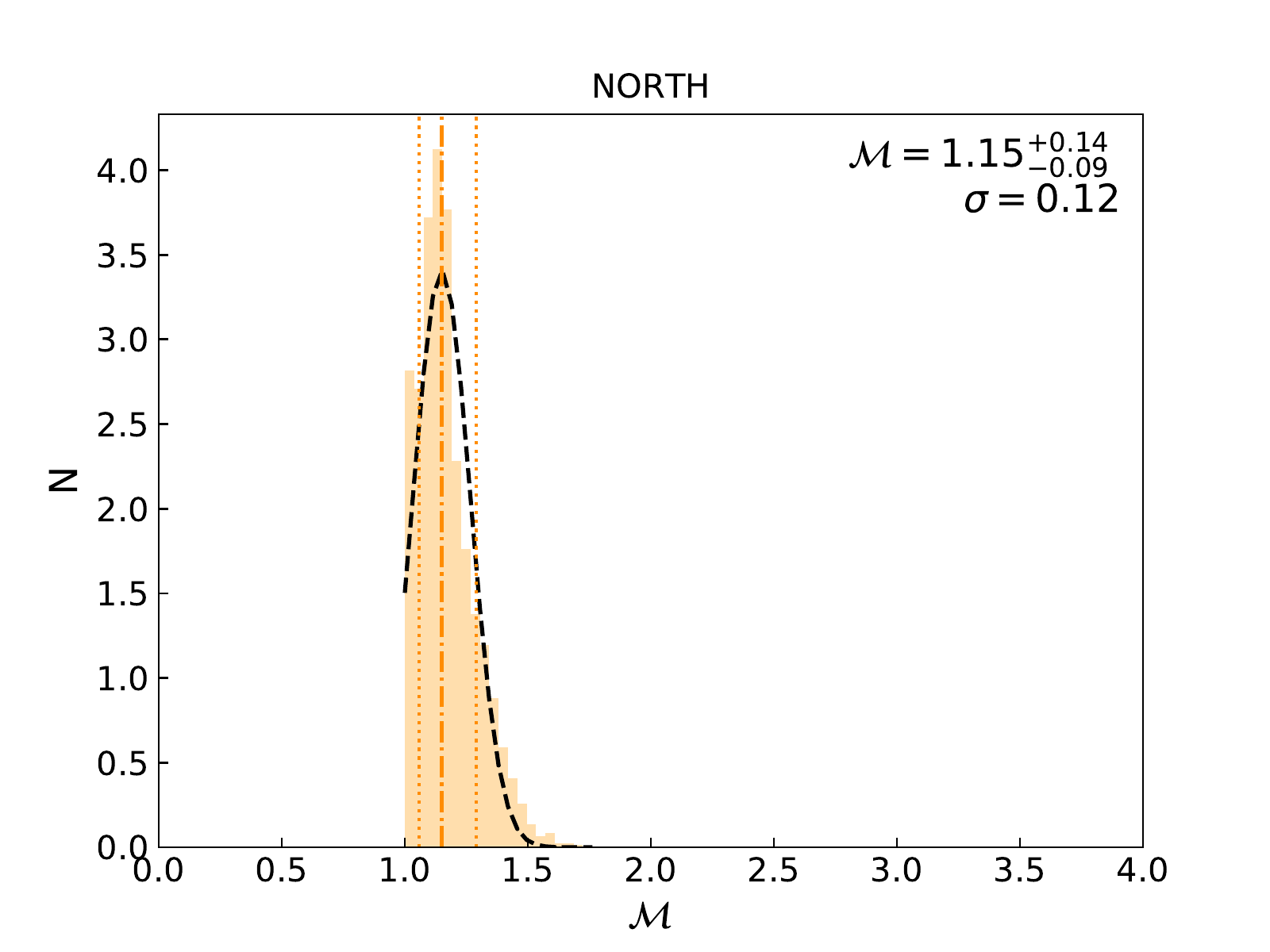}
}
\centerline{
\includegraphics[width=0.50\textwidth]{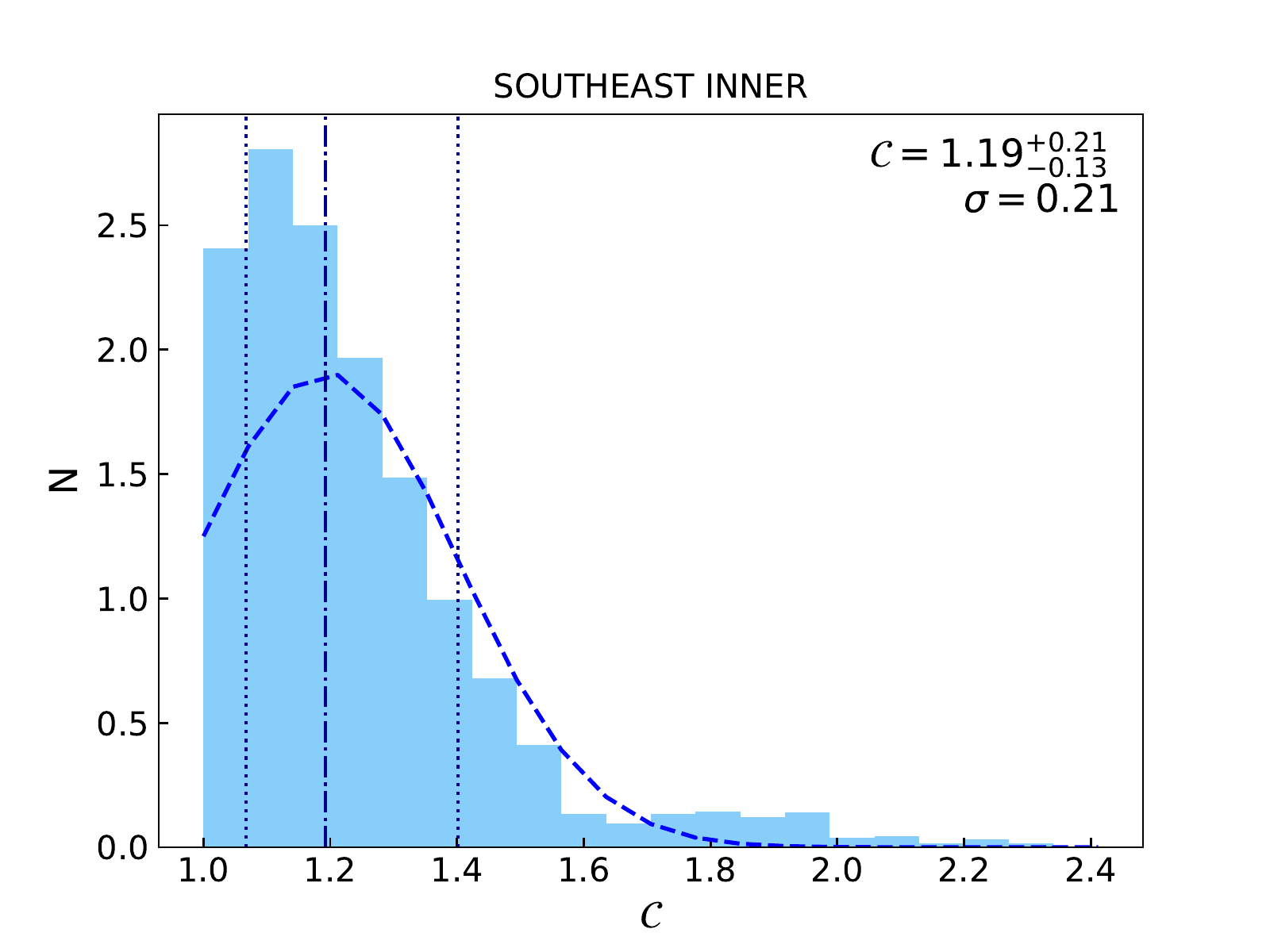}
\includegraphics[width=0.50\textwidth]{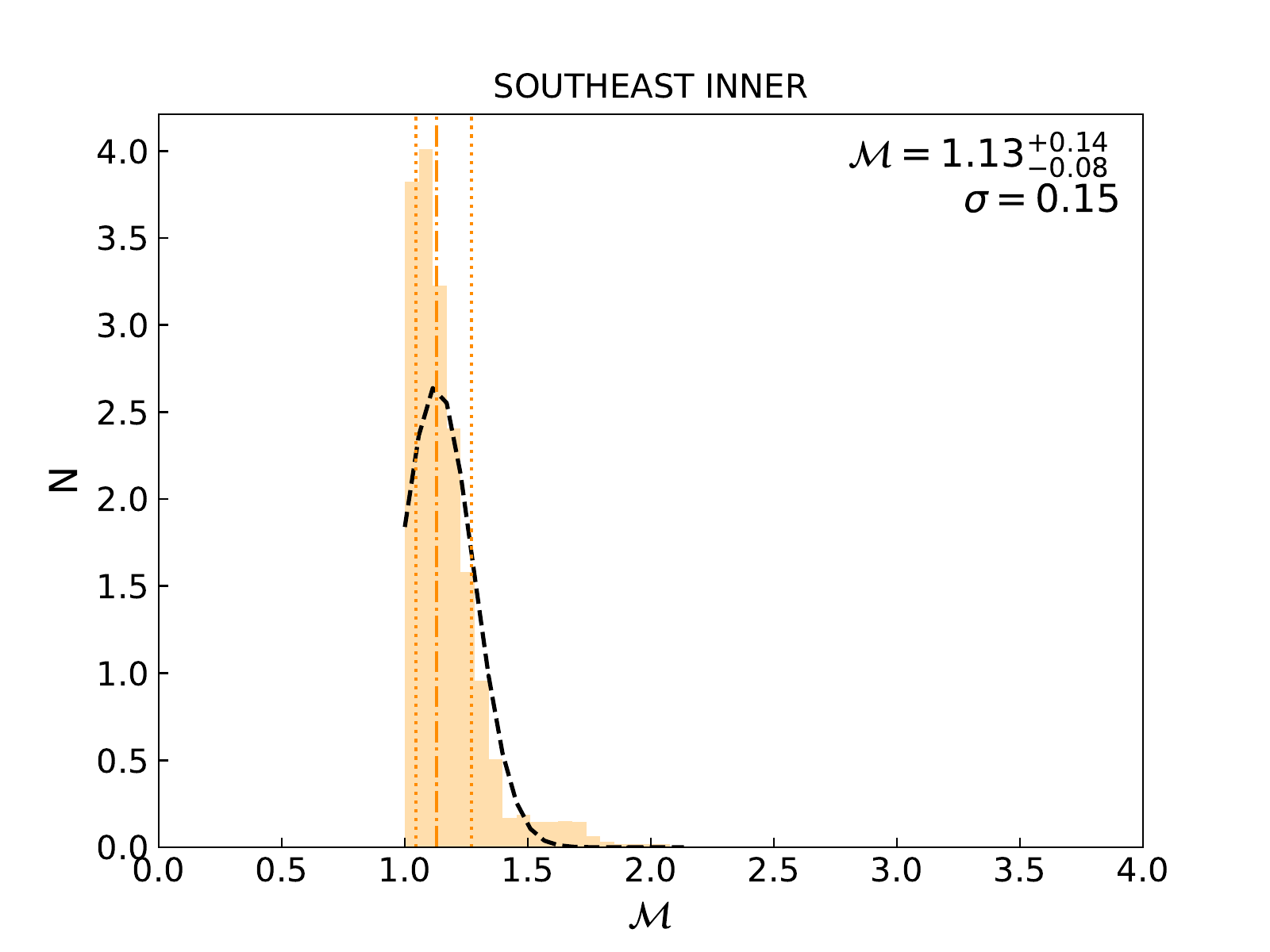}
}
\centerline{
\includegraphics[width=0.50\textwidth]{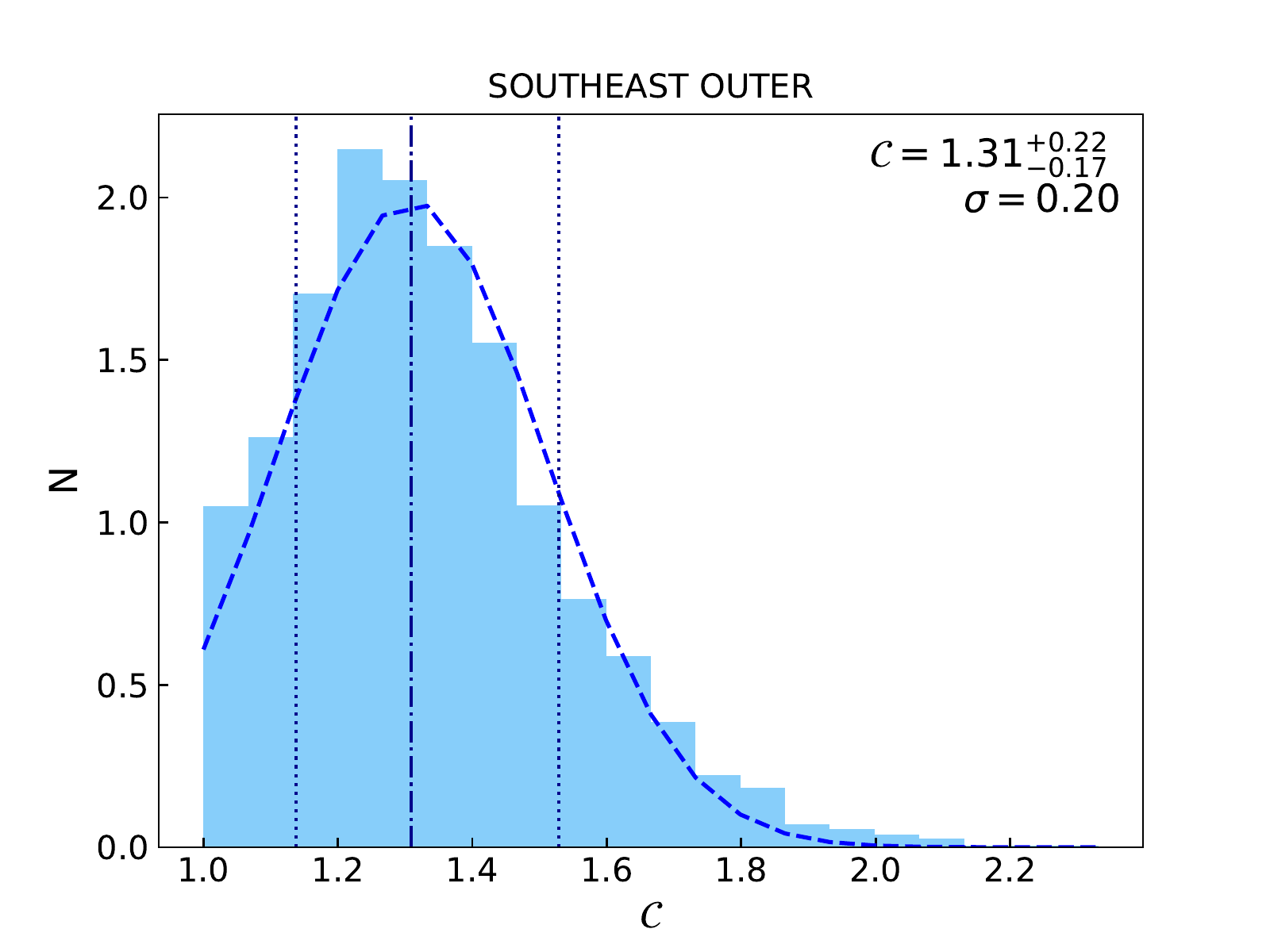}
\includegraphics[width=0.50\textwidth]{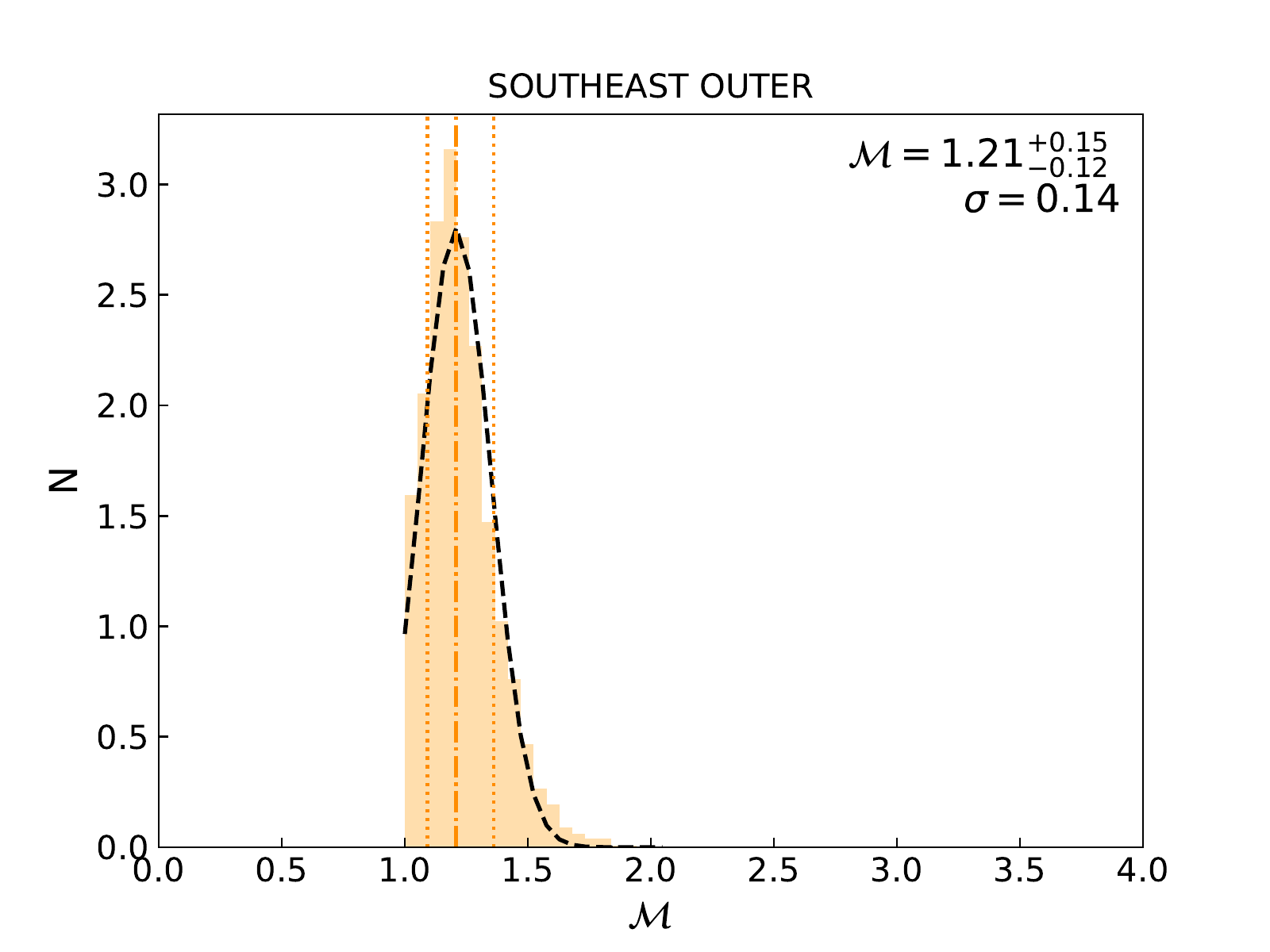}
}
\caption{\small{Distribution of density jumps (left panels) and Mach numbers (right panels) for the discontinuities in the norther and southern sectors. 
Top panels: distributions of density jumps and Mach numbers from the MCMC simulations from the discontinuity in the northern sector. Center panels: same as top panels, except for the southeast inner sector. Bottom panels: same as top panels, except for the southeast outer sector.
}}\label{fig:dist_north}
\end{figure*}

\newpage

\section{MCMC corner plots}\label{apdx:cornerplots}
Here we present the MCMC ``corner plots'' \citep{2016Foreman-Mackey,2017Foreman-Mackey} from Figures  \ref{fig:sb_northern}, \ref{fig:sb_south_east} and \ref{fig:sb_south}. They show the distribution of solutions for the fitted parameters from the X-ray surface brightness profiles across the wedges presented in those Figures. For all corner plots, contour levels are drawn at $[0.5, 1.0, 1.5, 2.0]\sigma$ levels.

\setcounter{figure}{0} 
\begin{figure*}[!b]
\centering
\includegraphics[width=1\textwidth]{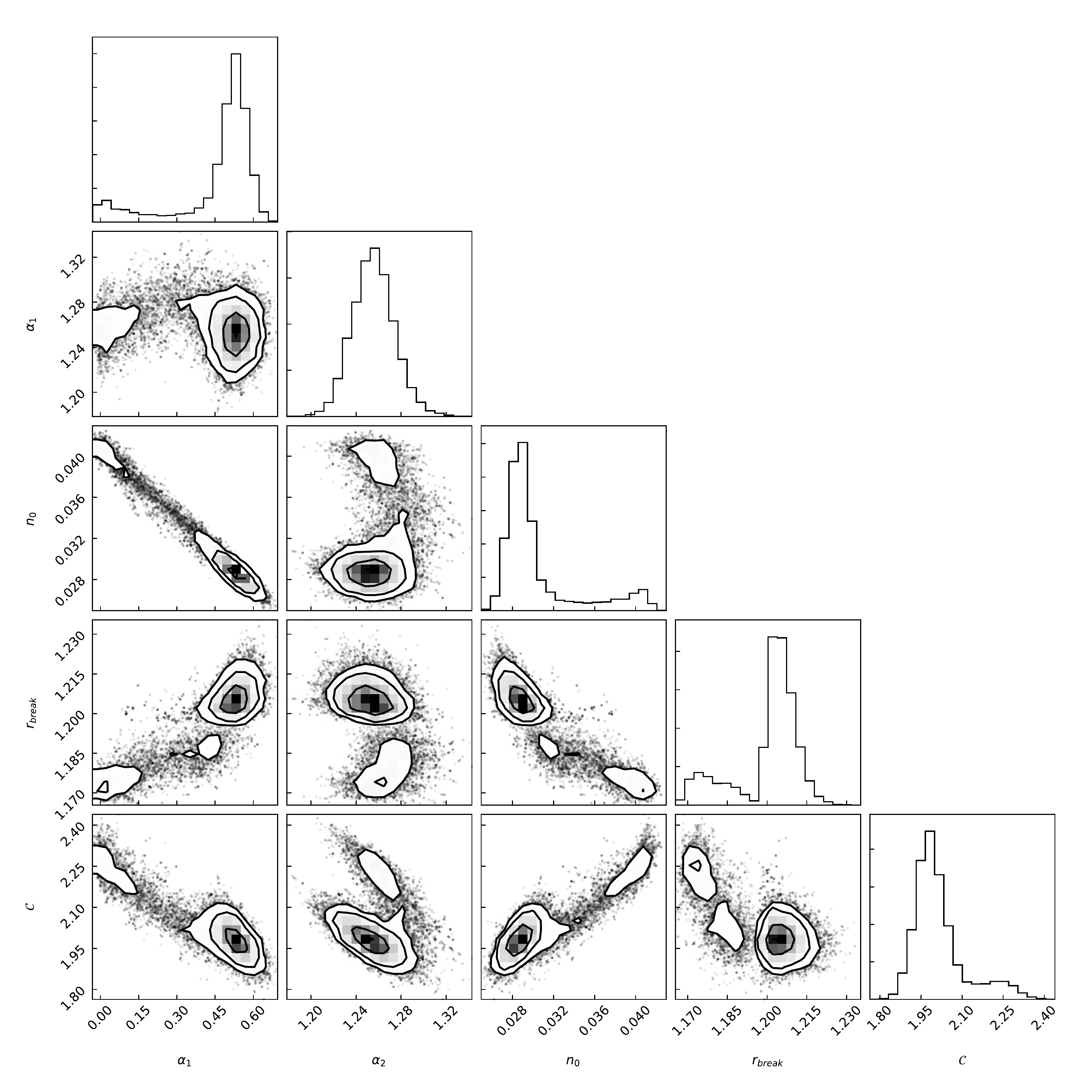}
\caption{The MCMC ``corner plot'' for the distribution of solutions of the fitted parameters from the X-ray surface brightness profiles across the bullet discontinuity (see left panel in Figure \ref{fig:sb_northern}).}\label{fig:corner_north_in}
\end{figure*}

\begin{figure*}
\centering
\includegraphics[width=1\textwidth]{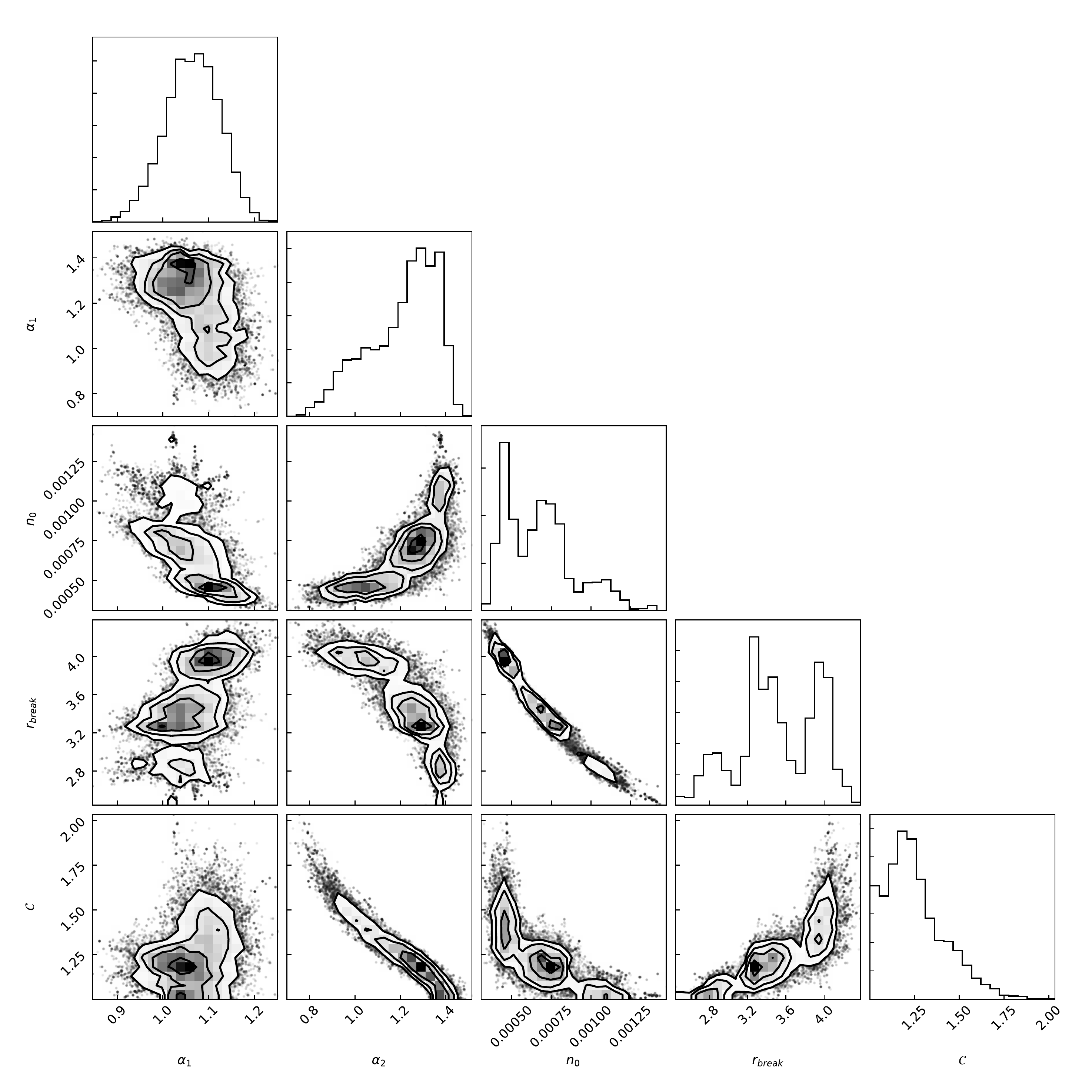}
\caption{The MCMC ``corner plot'' for the distribution of solutions of the fitted parameters from the X-ray surface brightness profiles across the northern discontinuity (see right panel of Figure \ref{fig:sb_northern}).}\label{fig:corner_north_out}
\end{figure*}

\begin{figure*}
\centering
\includegraphics[width=1\textwidth]{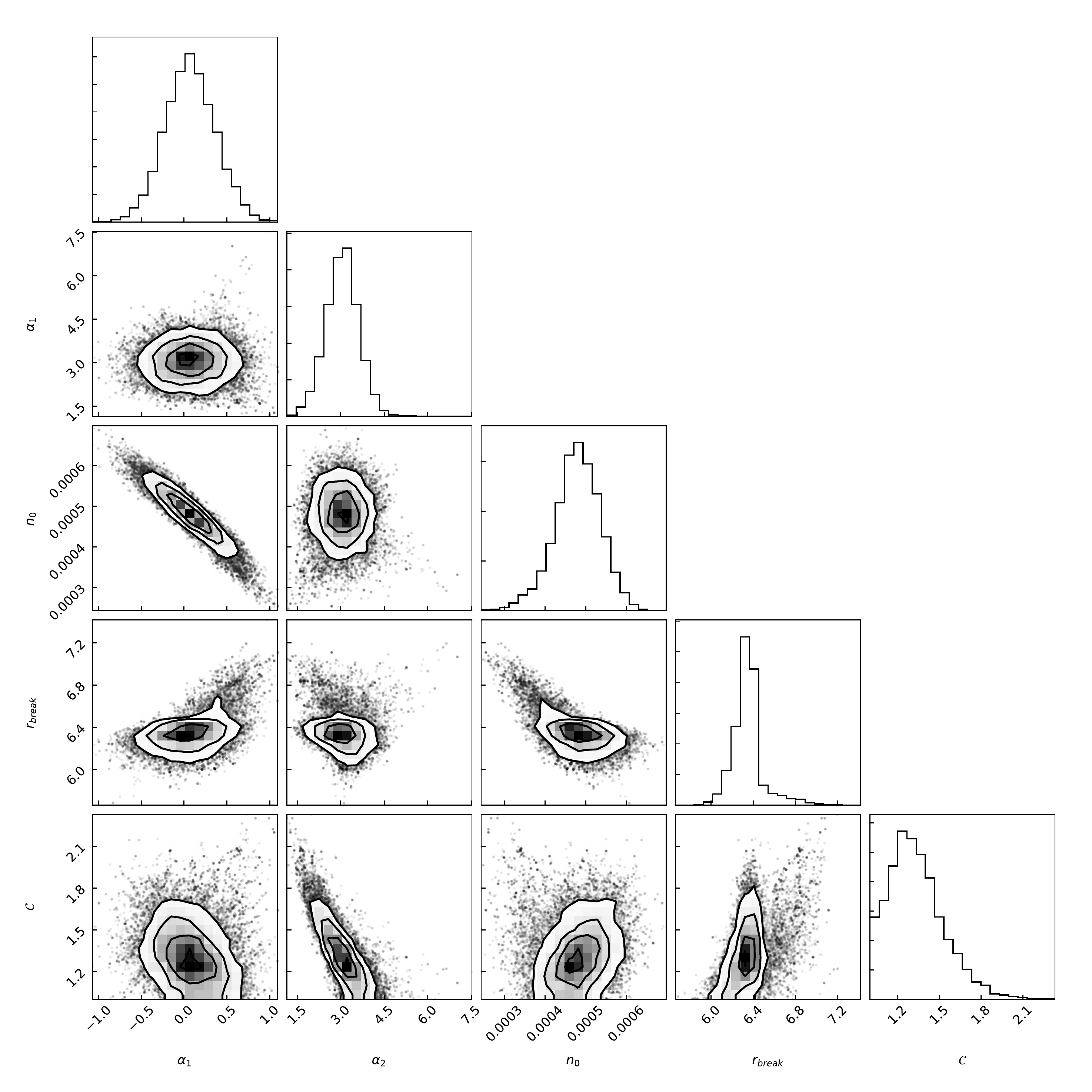}
\caption{The MCMC ``corner plot'' for the distribution of solutions of the fitted parameters from the X-ray surface brightness profiles across the discontinuity in the southeast outer wedge (see left panel in Figure \ref{fig:sb_south_east}).}\label{fig:corner_southeast_out}
\end{figure*}

\begin{figure*}
\centering
\includegraphics[width=1\textwidth]{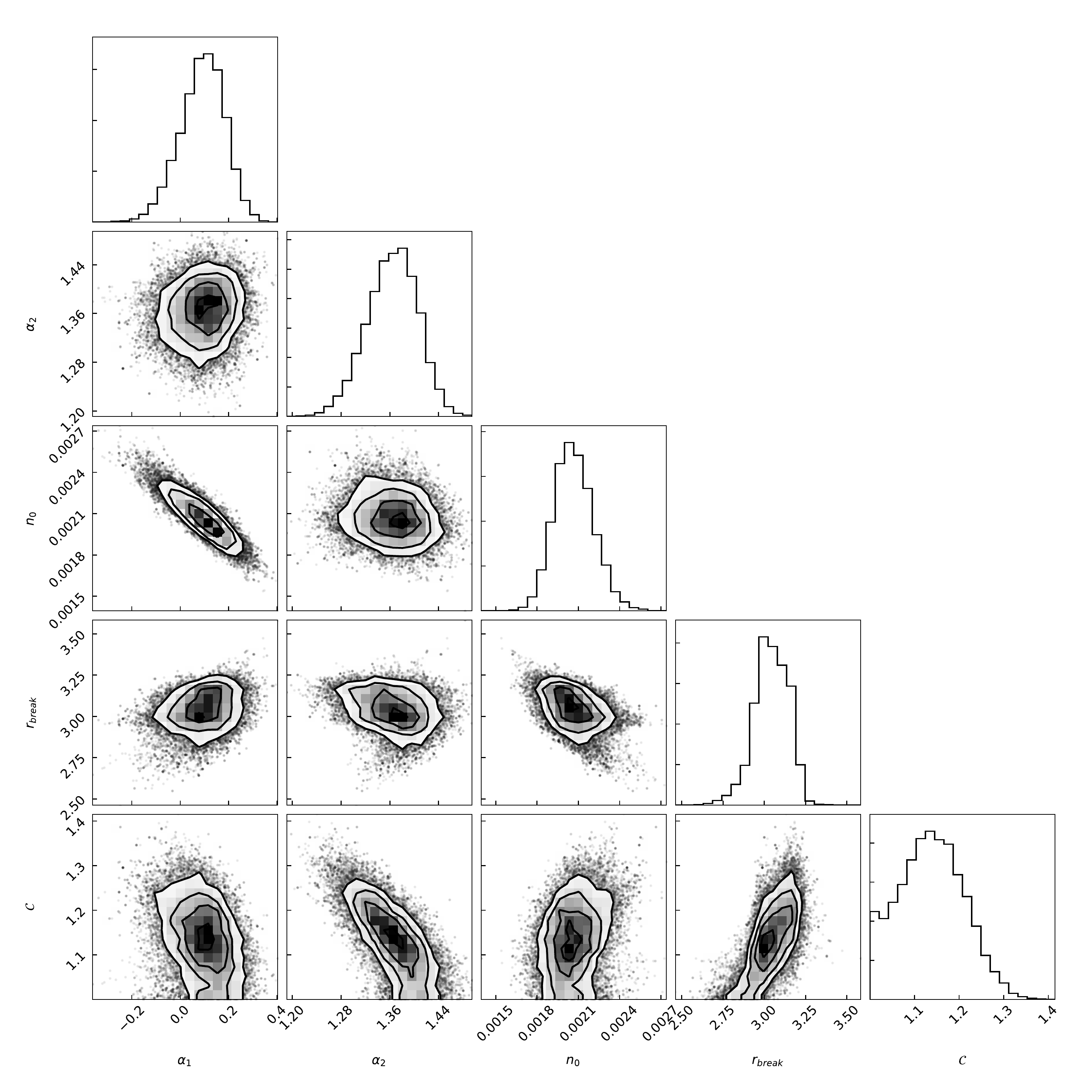}
\caption{The MCMC ``corner plot'' for the distribution of solutions of the fitted parameters from the X-ray surface brightness profiles across the discontinuity in the southeast inner wedge (see top right panel in Figure \ref{fig:sb_south_east}).}\label{fig:corner_southeast_in}
\end{figure*}

\begin{figure*}
\centering
\includegraphics[width=1\textwidth]{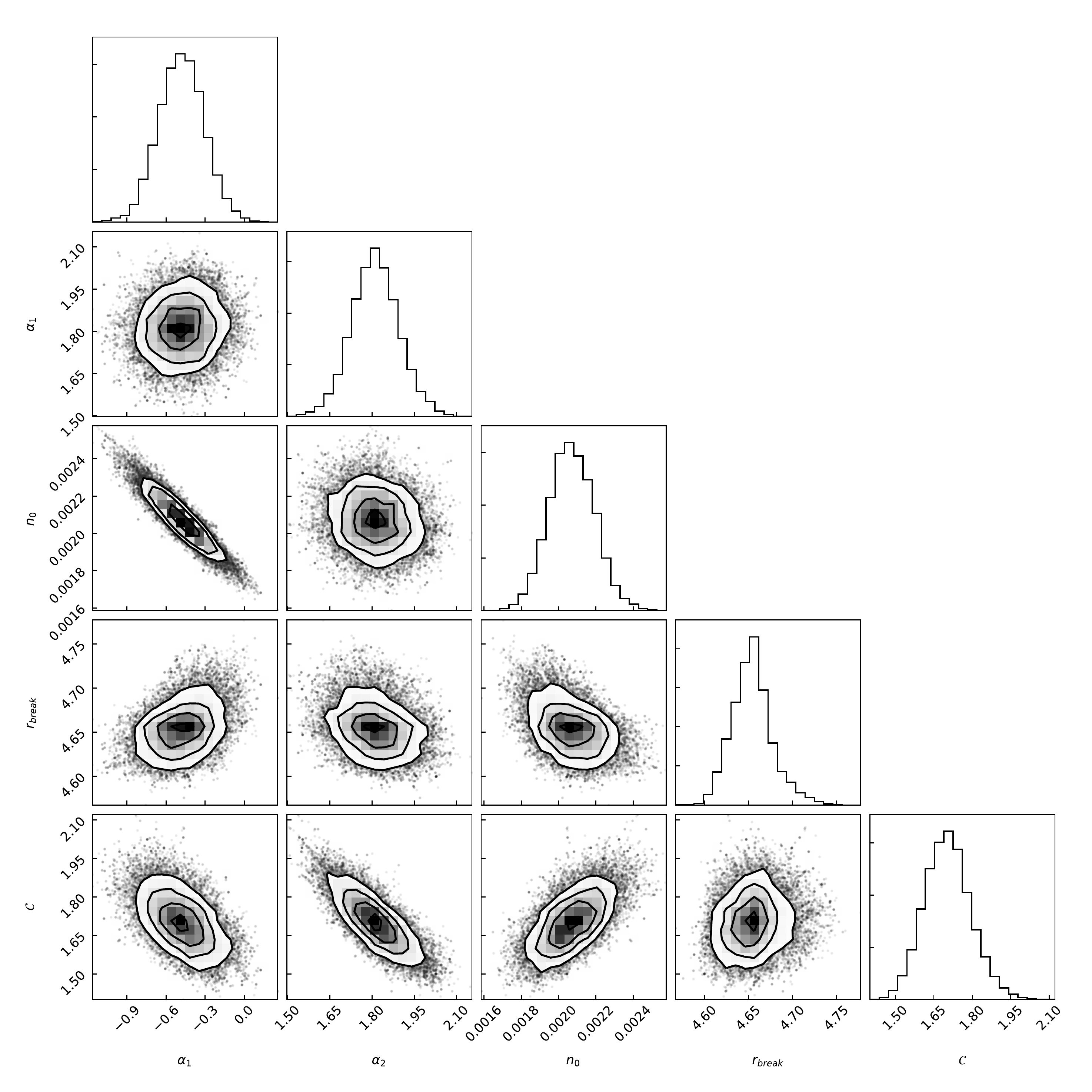}
\caption{The MCMC ``corner plot'' for the distribution of solutions of the fitted parameters from the X-ray surface brightness profiles across the discontinuity in the south wedge (see Figure \ref{fig:sb_south}).}\label{fig:corner_south}
\end{figure*}

\end{document}